%% file: main.tex
\documentclass[preprint,12pt]{elsarticle}

\usepackage{amssymb}
\usepackage{amsthm}
\usepackage{amsmath}
\usepackage{graphicx,graphics}

\usepackage{float}

\usepackage{algorithm}
\usepackage{algpseudocode}
\usepackage{multirow}

\usepackage{physics}
\usepackage{amsmath}
\usepackage{tikz}
\usepackage{mathdots}
\usepackage{yhmath}
\usepackage{cancel}
\usepackage{color}
\usepackage{siunitx}
\usepackage{array}
\usepackage{multirow}
\usepackage{amssymb}
\usepackage{gensymb}
\usepackage{tabularx}
\usepackage{extarrows}
\usepackage{booktabs}
\usetikzlibrary{fadings}
\usetikzlibrary{patterns}
\usetikzlibrary{shadows.blur}
\usetikzlibrary{shapes}

\usepackage{cancel}

\begin{document}

\begin{frontmatter}



\title{Development of a Novel Riemann Solver for Solid Dynamics}


\author[label1]{Khoder Alhamwi ALshaar}

\affiliation[label1]{organization={Aerospace Engineering Department, IIT Bombay},
            addressline={Powai}, 
            city={Mumbai},
            postcode={400076}, 
            state={Maharashtra},
            country={India}}
\ead{alshaar@iitb.ac.in}
\author[label1]{J C Mandal}
\ead{mandal@iitb.ac.in}

\input{sections/abstract}

\end{frontmatter}


\input{sections/introduction}
\input{sections/mathematical_model}

\input{sections/Numerical_formulation}
\input{sections/solution_algorithm}
\input{sections/test_cases/numerical_test_cases}
\input{sections/conclusion}

\newpage



 \bibliographystyle{elsarticle-num} 
 \bibliography{bib}





\end{document}

%% file: sections/abstract.tex
\begin{abstract}
This work presents a new finite volume framework for solid dynamics based on a momentum–deformation formulation. Building on the C-TOUCH methodology \cite{haider2018upwind}, a novel Roe-type Riemann solver is developed to enhance the stability and accuracy of hyperbolic conservation law solutions in solids. 
The approach naturally handles multidimensional problems and provides a foundation for future extensions to nonlinear and fluid–structure interaction cases. Validation against standard two- and three-dimensional linear elasticity benchmarks demonstrates the method’s robustness and accuracy relative to traditional displacement-based approaches, highlighting its promise for large-scale dynamic simulations.

\end{abstract}


%% file: sections/introduction.tex
\section{Introduction}
\label{sec:Introduction}

Recent advancements in computational methods for solid dynamics have significantly impacted various industrial applications, 
including crash simulations, impact analysis, and structural assessments. This paper presents a novel computational framework 
for solving linearly elastic solid dynamics problems, with future extensions toward nonlinear and fluid–solid interaction challenges.

Computational approaches for solid dynamics can be broadly classified based on the formulation: displacement-based \cite{demirdzic1988numerical,cardiff2016block,jasak2000application,reddy1993introduction}, momentum-defor-
mation-based \cite{trangenstein1991higher,lee2013development,haider2017first, bonet2021nonlinear,di2024first,lee2024novel}, momentum-stress-based  \cite{leveque2002finite,breil20203d,ortega2014numerical,li2022hllc,huerta1994new}. Each formulation offers a distinct perspective on the governing equations. Displacement-based formulations typically involve second-order differential equations, while momentum-deformation and momentum-stress-based formulations lead to first-order systems.

Another classification is based on the frame of reference, which may be Eulerian\cite{ortega2014numerical,li2022hllc} or Lagrangian (either Total \cite{trangenstein1991higher,lee2013development,haider2017first} or Updated  \cite{breil20203d,hassan2019upwind}), or Arbitrary Lagrangian-Eulerian (ALE) \cite{di2024first,huerta1994new,lee2024novel}. These approaches differ in how they associate the computational mesh with material motion: the Lagrangian frame aligns nodes with material points, the Eulerian frame fixes the mesh while allowing material to flow through it, and the ALE method combines both strategies.

Several numerical techniques are employed to implement these formulations, including the Finite Element Method (FEM), Finite Volume Method (FVM), and Smoothed Particle Hydrodynamics (SPH). FEM has traditionally dominated the field of solid dynamics, particularly in stress analysis \cite{jasak2000application}.
However, FVM has gained prominence due to its ability to produce diagonally dominant matrices suited for iterative solvers \cite{jasak2000application,cardiff2021thirty}, thereby improving computational efficiency for large-scale problems \cite{jasak2000application, trangenstein1991higher}. Unlike FEM, which relies on predefined shape functions and direct solvers, FVM offers greater flexibility for handling complex, nonlinear systems, extending its applications beyond fluid dynamics into solid mechanics.

In industrial contexts involving transient, large-strain deformations, explicit, low-order, displacement-based FEM formulations are widely used 
for their robustness and computational efficiency \cite{bonet2021nonlinear, lee2014development}. Nonetheless, these methods face challenges such as reduced strain and stress accuracy, 
difficulties with bending-dominated problems, volumetric and shear locking, and hydrostatic pressure oscillations \cite{bonet2021nonlinear,lee2014development,bonet2001stability,haider2017first}. Additionally, conventional 
time integration schemes, including Newmark-type methods, often introduce high-frequency noise near sharp spatial gradients, while artificial damping can further degrade accuracy \cite{hilber1977improved,lee2014development, bonet2021nonlinear}. These issues limit the suitability of traditional schemes for wave propagation and shock-dominated 
problems \cite{bonet2021nonlinear, abd2014two, haider2017first}.

First-order system formulations, commonly used in computational fluid dynamics and electromagnetics, provide a more natural and stable representation of physical processes \cite{leveque2002finite}. When discretized using the finite volume method, these systems improve convergence and ensure consistent accuracy. In line with these advantages, we adopt the momentum-deformation-based formulation in this work, as it offers several benefits over traditional displacement-based approaches \cite{bonet2021nonlinear, cardiff2021thirty, lee2013development, haider2019toolkit}.

P. Cardiff and I. Demirdžić \cite{cardiff2021thirty} provide a comprehensive review of the FVM developments in solid mechanics over the past three decades, highlighting key differences from FEM. Early work by Trangenstein and Colella \cite{trangenstein1991higher} introduced a momentum–deformation-based FVM approach, which has since evolved through contributions involving FEM, FVM, and SPH methodologies \cite{haider2018upwind}. A notable contribution to this evolution is the upwind, cell-centered FVM framework developed by Lee et al. \cite{lee2013development}, specifically tailored for efficient transient dynamic simulations involving large deformations. Building upon this, Haider et al. \cite{haider2017first, haider2018upwind, haider2019toolkit} further extended the approach within the OpenFOAM V.7 platform by developing the \verb|explicitSolidsDynamics| toolkit, a cell centered explicit solid dynamics solver for OpenFOAM \cite{haider2019toolkit}, culminating in the formulation of the C-TOUCH (Constrained-Total Lagrangian Upwind Cell-Centered Finite Volume Method for Hyperbolic Conservation Laws) framework.

C-TOUCH incorporates advanced techniques such as second-order spatial reconstruction, an acoustic Riemann solver based on approximate-state Riemann solvers \cite{toro2013riemann}, and a two-stage, single-time Strong Stability Preserving Runge–Kutta (SSPRK) time integration scheme. It also enforces the curl-free condition on the deformation gradient using a constrained FVM strategy similar to that employed in magnetohydrodynamics \cite{toth2000b, torrilhon2005locally}, and ensures conservation of angular momentum through a discrete projection algorithm.

Building upon C-TOUCH, the present work introduces a new Roe-type Riemann solver tailored for solid dynamics applications. Developed within the OpenFOAM V.7 platform by extending the \verb|expllicitSolidsDynamics| toolkit, the solver integrates seamlessly with the existing finite volume infrastructure. Riemann solvers are recognized as powerful tools for solving hyperbolic conservation laws \cite{leveque2002finite}, and the momentum–deformation-based formulation,  being a first-order system with hyperbolicity, homogeneity, and rotational invariance, provides a robust mathematical foundation for their development \cite{toro2013riemann}. The newly developed Roe-type solver offers a simple yet effective means for addressing multidimensional problems and holds promise for future extensions to nonlinear elasticity and fluid–structure interaction problems.

To validate the proposed method, several two and three-dimensional linear elasticity benchmark problems are solved, including shock wave propagation, stress analysis, and dynamic deformation. Results are compared against analytical and numerical solutions, demonstrating the accuracy and robustness of the new solver. 

 The remainder of this paper is organized as follows: Section \ref{sec:Governing Equations} presents the Governing Equations based on momentum-deformation-based formulation and the linear elasticity constitutive law. Section \ref{sec:Numerical Discretization} outlines the numerical discretization procedure, including the finite volume method (FVM), time integration using the SSPRK scheme, numerical flux evaluation with the Roe-type Riemann solver, and the treatment of boundary conditions. Section \ref{sec:Solution Algorithm} describes the overall solution algorithm,  including constraint enforcement and conservation of angular momentum. Section \ref{sec:Numerical Test Cases} provides numerical examples and validation studies. Finally, section \ref{sec:Conclusion} concludes the paper with final remarks and directions for future research.

%% file: sections/mathematical_model.tex
\section{Governing Equations}
\label{sec:Governing Equations}
The conservation equations for linear momentum $\boldsymbol{p}$ and the deformation gradient tensor $\boldsymbol{F}$ in the Lagrangian frame are given in differential form as follows:
\begin{equation} 
\begin{aligned}
	\frac{\partial \boldsymbol{p}}{\partial t} - \frac{\partial \boldsymbol{(P E_I)}}{\partial \boldsymbol{X_I}} = \rho_0 \boldsymbol{b}, \\
\frac{\partial \boldsymbol{F}}{\partial t} - \frac{\partial (\frac{1}{\rho_0} \boldsymbol{p} \otimes \boldsymbol{E}_I)}{\partial \boldsymbol{X_I}} = 0,  
\end{aligned}
\qquad \forall\, I = 1,2,3.
\label{eq: governing equations}
\end{equation}
where $\boldsymbol{P}$ is the first Piola-Kirchhoff stress tensor, $\rho_0$ is the (reference) material density, $\boldsymbol{b}$ is the body forces per unit mass, $\boldsymbol{E}_I$ denotes the Cartesian basis vector in the $I^{th}$ material direction. These equations describe conservation of momentum and deformation evolution, respectively.

Equation~\eqref{eq: governing equations} can be recast into a compact, first-order hyperbolic system of the form:
\begin{equation}
	\frac{\partial\boldsymbol{\mathcal{U}}}{\partial t} + \frac{\partial\boldsymbol{\mathcal{F}}_{1}}{\partial {X}_{2}} + \frac{\partial\boldsymbol{\mathcal{F}}_{2}}{\partial {X}_{2}} + \frac{\partial\boldsymbol{\mathcal{F}}_{3}}{\partial {X}_{3}} = \boldsymbol{\mathcal{S}},
    \label{eq:first-order system}
\end{equation} 
where $\boldsymbol{\mathcal{U}}$ is the vector of conservative variables, $\boldsymbol{\mathcal{F}}_I$ is the flux vector in the $I^{th}$ direction, and $\boldsymbol{\mathcal{S}}$ is the source term vector. 

The system components are defined as:
\begin{equation}
	\boldsymbol{\mathcal{U}}=
	\begin{bmatrix}
		p_1\\
		p_2\\
		p_3\\
		F_{11}\\
		F_{12}\\
		F_{13}\\
		F_{21}\\
		F_{22}\\
		F_{23}\\
		F_{31}\\
		F_{32}\\
		F_{33}
	\end{bmatrix}
	,\,\,
	\boldsymbol{\mathcal{F}_1}= -
	\begin{bmatrix}
		{P}_{11}\\
		{P}_{21}\\
		{P}_{31}\\
		\frac{1}{\rho }{p}_1\\
		0\\
		0\\
		\frac{1}{\rho }{p}_2\\
		0\\
		0\\
		\frac{1}{\rho }{p}_3\\
		0\\
		0
	\end{bmatrix}
	,\,\,
	\boldsymbol{{\mathcal{F}}_2}= -
	\begin{bmatrix}
		{P}_{12}\\
		{P}_{22}\\
		{P}_{32}\\
		0\\
		\frac{1}{\rho }{p}_1\\
		0\\
		0\\
		\frac{1}{\rho }{p}_2\\
		0\\
		0\\
		\frac{1}{\rho }{p}_3\\
		0
	\end{bmatrix}
	,\,\,
	\boldsymbol{\mathcal{F}_3}= -
	\begin{bmatrix}
		{P}_{13}\\
		{P}_{23}\\
		{P}_{33}\\
		0\\
		0\\
		\frac{1}{\rho }{p}_1\\
		0\\
		0\\
		\frac{1}{\rho }{p}_2\\
		0\\
		0\\
		\frac{1}{\rho }{p}_3
	\end{bmatrix}
	, \,\, \boldsymbol{S} = 
	\begin{bmatrix}
		{b}_{1}\\
		{b}_{2}\\
		{b}_{3}\\
		0\\
		0\\
		0\\
		0\\
		0\\
		0\\
		0\\
		0\\
		0
	\end{bmatrix}
\end{equation}

\noindent To close the system, a constitutive relation is required to link the stress tensor $\boldsymbol{P}$ with the deformation gradient $\boldsymbol{{F}}$. For a linearly elastic solid, the constitutive law is given by:
\begin{equation}
P(\boldsymbol{F})=\mu\left[\boldsymbol{F}+\boldsymbol{F}^T-\frac{2}{3}\mathrm{tr}(\boldsymbol{F})\boldsymbol{I}\right]+(\lambda+\frac23 \mu)\left(\mathrm{tr}(\boldsymbol{F})-3\right)\boldsymbol{I},
\end{equation}
where $\lambda$ and $\mu$ are Lame's parameters defined in terms of the material's Young's modulus $E$ and Poisson's ratio $\nu$ as:
 \begin{equation}
     \lambda = \frac{\nu E}{(1+\nu)(1-2\nu)}, \quad \mu = \frac{E}{2(1+\nu)}.
 \end{equation}

%% file: sections/Numerical_formulation.tex
\section{Numerical Method}
\label{sec:Numerical Discretization}
In this paper, we discretize the system given in equations~\eqref{eq:first-order system} using a cell-centered finite volume method based on the method of lines, where spatial and temporal discretizations are performed separately. The flux vector is then computed using a novel Roe-type Riemann solver combined with a second-order reconstruction technique. Simultaneously, the time integration is carried out using a two-stage Runge–Kutta scheme.

\subsection{Finite Volume Discretization}
\label{subsec: finite volume discretization}
\input{figures/finite_volume_cell}
In the finite volume method, the integral form of the governing equations~\eqref{eq:first-order system} is employed and can be expressed as:
\begin{equation}
    {\frac{d}{d t}}\int_{\Omega}\, \boldsymbol{ \cal U}\,d\Omega =\int_{\partial\Omega} (\boldsymbol{ \cal F}_1, \boldsymbol{ \cal F}_2, \boldsymbol{ \cal F}_3) \cdot \boldsymbol{N}\,d S + \int_{\Omega}\, \boldsymbol{ \cal S}\,d\Omega.
    \label{eq:integral form first-order system}
\end{equation}
Here, $\Omega$ denotes an arbitrary control volume, $\partial \Omega$ is its boundary surface, $(\boldsymbol{ \cal F}_1, \boldsymbol{ \cal F}_2, \boldsymbol{ \cal F}_3)$ is the flux tensor, and $\boldsymbol{N}= [N_1, N_1, N_3]^T$ is the outward-pointing unit normal vector on the surface. 

The semi-discrete form of the Eq.~\eqref{eq:integral form first-order system} for the finite volume cell $e$ (as illustrated in Fig.\ref{fig:finite volume cell }) after space discretization can be written as:

\begin{equation}
   \frac{\mathrm{d} \boldsymbol{\cal{U}}_e}{\mathrm{d} t}=  = -\frac{1}{\Omega_e }\sum_{ f \in {\Lambda_e^f}} (\boldsymbol{ \cal F}_1, \boldsymbol{ \cal F}_2, \boldsymbol{ \cal F}_3)_f \cdot \boldsymbol{N}_f \, \Delta S_f+  \boldsymbol{\cal{S}}_e,
   \label{eq:semi-discretized}
\end{equation}
where $\boldsymbol{\cal{U}}_e$ is is the cell-averaged conserved variable vector, $\Lambda_e^f$ is the number of faces of the control volume, $\boldsymbol{N}_f$ is the outward unit normal vector on face $f$, and $\Delta S_f$ is the corresponding face area. The flux tensor $(\boldsymbol{ \cal F}_1, \boldsymbol{ \cal F}_2, \boldsymbol{ \cal F}_3)_f$ is evaluated at the mid-point of each cell interface. $\boldsymbol{\cal{S}}_e$ is the cell-averaged source vector.

This spatial discretization reduces the original system of first-order hyperbolic partial differential equations Eq.~\eqref{eq:first-order system} to a set of ordinary differential equations. Upon expanding Eq.~\eqref{eq:semi-discretized}, the evolution equations for linear momentum $\boldsymbol{p}$ and the deformation gradient tensor $\boldsymbol{F}$ can be expressed as:
\begin{subequations}
\begin{align}
        \frac{\mathrm{d} \boldsymbol{p}_e}{\mathrm{d} t} =  - \frac{1}{\Omega_e }  \sum_{f=1}^M (\boldsymbol{{P}} \boldsymbol{N} \, \Delta S)_f +  \rho \boldsymbol{b}_e \label{eq : semi-discret linear momentum},\\
        \frac{\mathrm{d} \boldsymbol{F}_e}{\mathrm{d} t} =  - \frac{1}{\Omega_e }  \sum_{f=1}^M (\frac{1}{\rho}\boldsymbol{{p}} \otimes\boldsymbol{N} \, \Delta S)_f.
        \label{eq : semi-discret deformation}
  \end{align}
\end{subequations}

\input{sections/numerical_flux_evaluation}

\input{sections/time-integration}
\input{sections/boundary_conditions}

%% file: figures/finite_volume_cell.tex
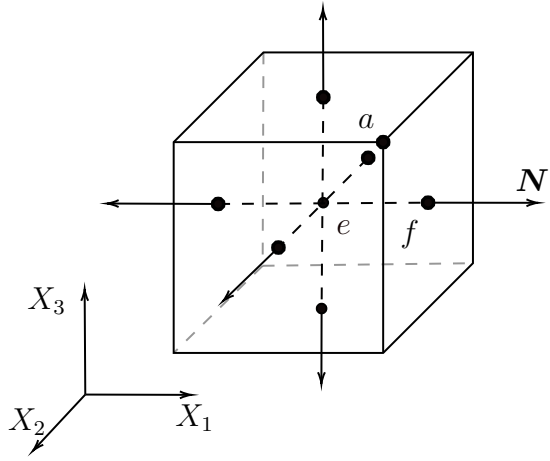
\begin{figure}[t]
    \centering
\tikzset{every picture/.style={line width=0.75pt}} 

\begin{tikzpicture}[x=0.4pt,y=0.4pt,yscale=-1,xscale=1]

\draw   (161,137.5) -- (245.9,52.6) -- (444,52.6) -- (444,252.1) -- (359.1,337) -- (161,337) -- cycle ; \draw   (444,52.6) -- (359.1,137.5) -- (161,137.5) ; \draw   (359.1,137.5) -- (359.1,337) ;
\draw [color={rgb, 255:red, 0; green, 0; blue, 0 }  ,draw opacity=0.41 ] [dash pattern={on 4.5pt off 4.5pt}]  (245.13,254.53) -- (444,252.1) ;
\draw [color={rgb, 255:red, 0; green, 0; blue, 0 }  ,draw opacity=0.41 ] [dash pattern={on 4.5pt off 4.5pt}]  (245.13,254.53) -- (161,337) ;
\draw [color={rgb, 255:red, 0; green, 0; blue, 0 }  ,draw opacity=0.41 ] [dash pattern={on 4.5pt off 4.5pt}]  (245.9,52.6) -- (245.13,254.53) ;
\draw  [fill={rgb, 255:red, 15; green, 8; blue, 8 }  ,fill opacity=1 ][line width=0.75]  (297.97,194.8) .. controls (297.97,192.3) and (300,190.27) .. (302.5,190.27) .. controls (305,190.27) and (307.03,192.3) .. (307.03,194.8) .. controls (307.03,197.3) and (305,199.33) .. (302.5,199.33) .. controls (300,199.33) and (297.97,197.3) .. (297.97,194.8) -- cycle ;
\draw    (401.55,194.8) -- (503.53,194.93) ;
\draw [shift={(505.53,194.93)}, rotate = 180.07] [color={rgb, 255:red, 0; green, 0; blue, 0 }  ][line width=0.75]    (10.93,-3.29) .. controls (6.95,-1.4) and (3.31,-0.3) .. (0,0) .. controls (3.31,0.3) and (6.95,1.4) .. (10.93,3.29)   ;
\draw  [line width=3.75]  (399.35,194.8) .. controls (399.35,193.58) and (400.33,192.6) .. (401.55,192.6) .. controls (402.77,192.6) and (403.75,193.58) .. (403.75,194.8) .. controls (403.75,196.02) and (402.77,197) .. (401.55,197) .. controls (400.33,197) and (399.35,196.02) .. (399.35,194.8) -- cycle ;
\draw  [line width=3.75]  (356.9,137.5) .. controls (356.9,136.28) and (357.88,135.3) .. (359.1,135.3) .. controls (360.32,135.3) and (361.3,136.28) .. (361.3,137.5) .. controls (361.3,138.72) and (360.32,139.7) .. (359.1,139.7) .. controls (357.88,139.7) and (356.9,138.72) .. (356.9,137.5) -- cycle ;
\draw  [line width=3.75]  (200.87,196.02) .. controls (200.87,194.8) and (201.85,193.82) .. (203.07,193.82) .. controls (204.28,193.82) and (205.27,194.8) .. (205.27,196.02) .. controls (205.27,197.23) and (204.28,198.22) .. (203.07,198.22) .. controls (201.85,198.22) and (200.87,197.23) .. (200.87,196.02) -- cycle ;
\draw  [dash pattern={on 4.5pt off 4.5pt}]  (203.07,196.02) -- (306.19,195.38) -- (306.66,195.38) -- (401.55,194.8) ;
\draw  [dash pattern={on 4.5pt off 4.5pt}]  (302.5,95.05) -- (300.88,294.98) ;
\draw  [line width=3.75]  (300.3,95.05) .. controls (300.3,93.83) and (301.28,92.85) .. (302.5,92.85) .. controls (303.72,92.85) and (304.7,93.83) .. (304.7,95.05) .. controls (304.7,96.27) and (303.72,97.25) .. (302.5,97.25) .. controls (301.28,97.25) and (300.3,96.27) .. (300.3,95.05) -- cycle ;
\draw  [line width=2.25]  (298.68,294.98) .. controls (298.68,293.77) and (299.66,292.78) .. (300.88,292.78) .. controls (302.09,292.78) and (303.08,293.77) .. (303.08,294.98) .. controls (303.08,296.2) and (302.09,297.18) .. (300.88,297.18) .. controls (299.66,297.18) and (298.68,296.2) .. (298.68,294.98) -- cycle ;
\draw    (203.07,196.02) -- (100,195.61) ;
\draw [shift={(98,195.6)}, rotate = 0.23] [color={rgb, 255:red, 0; green, 0; blue, 0 }  ][line width=0.75]    (10.93,-3.29) .. controls (6.95,-1.4) and (3.31,-0.3) .. (0,0) .. controls (3.31,0.3) and (6.95,1.4) .. (10.93,3.29)   ;
\draw    (300.88,294.98) -- (300.7,364.25) ;
\draw [shift={(300.7,366.25)}, rotate = 270.14] [color={rgb, 255:red, 0; green, 0; blue, 0 }  ][line width=0.75]    (10.93,-3.29) .. controls (6.95,-1.4) and (3.31,-0.3) .. (0,0) .. controls (3.31,0.3) and (6.95,1.4) .. (10.93,3.29)   ;
\draw    (302.5,95.05) -- (302.21,13.07) ;
\draw [shift={(302.2,11.07)}, rotate = 89.8] [color={rgb, 255:red, 0; green, 0; blue, 0 }  ][line width=0.75]    (10.93,-3.29) .. controls (6.95,-1.4) and (3.31,-0.3) .. (0,0) .. controls (3.31,0.3) and (6.95,1.4) .. (10.93,3.29)   ;
\draw  [line width=3.75]  (342.75,152.35) .. controls (342.75,151.13) and (343.73,150.15) .. (344.95,150.15) .. controls (346.17,150.15) and (347.15,151.13) .. (347.15,152.35) .. controls (347.15,153.57) and (346.17,154.55) .. (344.95,154.55) .. controls (343.73,154.55) and (342.75,153.57) .. (342.75,152.35) -- cycle ;
\draw  [line width=3.75]  (257.85,237.25) .. controls (257.85,236.03) and (258.83,235.05) .. (260.05,235.05) .. controls (261.27,235.05) and (262.25,236.03) .. (262.25,237.25) .. controls (262.25,238.47) and (261.27,239.45) .. (260.05,239.45) .. controls (258.83,239.45) and (257.85,238.47) .. (257.85,237.25) -- cycle ;
\draw    (260.05,237.25) -- (209.08,286.53) ;
\draw [shift={(207.64,287.92)}, rotate = 315.97] [color={rgb, 255:red, 0; green, 0; blue, 0 }  ][line width=0.75]    (10.93,-3.29) .. controls (6.95,-1.4) and (3.31,-0.3) .. (0,0) .. controls (3.31,0.3) and (6.95,1.4) .. (10.93,3.29)   ;
\draw  [dash pattern={on 4.5pt off 4.5pt}]  (344.95,152.35) -- (260.05,237.25) ;
\draw    (77.4,376.4) -- (76.81,278.7) ;
\draw [shift={(76.8,276.7)}, rotate = 89.66] [color={rgb, 255:red, 0; green, 0; blue, 0 }  ][line width=0.75]    (10.93,-3.29) .. controls (6.95,-1.4) and (3.31,-0.3) .. (0,0) .. controls (3.31,0.3) and (6.95,1.4) .. (10.93,3.29)   ;
\draw    (77.4,376.4) -- (175.8,376.69) ;
\draw [shift={(177.8,376.7)}, rotate = 180.17] [color={rgb, 255:red, 0; green, 0; blue, 0 }  ][line width=0.75]    (10.93,-3.29) .. controls (6.95,-1.4) and (3.31,-0.3) .. (0,0) .. controls (3.31,0.3) and (6.95,1.4) .. (10.93,3.29)   ;
\draw    (77.4,376.4) -- (29.16,428.24) ;
\draw [shift={(27.8,429.7)}, rotate = 312.94] [color={rgb, 255:red, 0; green, 0; blue, 0 }  ][line width=0.75]    (10.93,-3.29) .. controls (6.95,-1.4) and (3.31,-0.3) .. (0,0) .. controls (3.31,0.3) and (6.95,1.4) .. (10.93,3.29)   ;
\draw  [fill={rgb, 255:red, 15; green, 8; blue, 8 }  ,fill opacity=1 ][line width=0.75]  (397.02,194.8) .. controls (397.02,192.3) and (399.05,190.27) .. (401.55,190.27) .. controls (404.05,190.27) and (406.08,192.3) .. (406.08,194.8) .. controls (406.08,197.3) and (404.05,199.33) .. (401.55,199.33) .. controls (399.05,199.33) and (397.02,197.3) .. (397.02,194.8) -- cycle ;
\draw  [fill={rgb, 255:red, 15; green, 8; blue, 8 }  ,fill opacity=1 ][line width=0.75]  (340.42,152.35) .. controls (340.42,149.85) and (342.45,147.82) .. (344.95,147.82) .. controls (347.45,147.82) and (349.48,149.85) .. (349.48,152.35) .. controls (349.48,154.85) and (347.45,156.88) .. (344.95,156.88) .. controls (342.45,156.88) and (340.42,154.85) .. (340.42,152.35) -- cycle ;
\draw  [fill={rgb, 255:red, 15; green, 8; blue, 8 }  ,fill opacity=1 ][line width=0.75]  (354.57,137.5) .. controls (354.57,135) and (356.6,132.97) .. (359.1,132.97) .. controls (361.6,132.97) and (363.63,135) .. (363.63,137.5) .. controls (363.63,140) and (361.6,142.03) .. (359.1,142.03) .. controls (356.6,142.03) and (354.57,140) .. (354.57,137.5) -- cycle ;
\draw  [fill={rgb, 255:red, 15; green, 8; blue, 8 }  ,fill opacity=1 ][line width=0.75]  (198.53,196.02) .. controls (198.53,193.51) and (200.56,191.48) .. (203.07,191.48) .. controls (205.57,191.48) and (207.6,193.51) .. (207.6,196.02) .. controls (207.6,198.52) and (205.57,200.55) .. (203.07,200.55) .. controls (200.56,200.55) and (198.53,198.52) .. (198.53,196.02) -- cycle ;
\draw  [fill={rgb, 255:red, 15; green, 8; blue, 8 }  ,fill opacity=1 ][line width=0.75]  (255.52,237.25) .. controls (255.52,234.75) and (257.55,232.72) .. (260.05,232.72) .. controls (262.55,232.72) and (264.58,234.75) .. (264.58,237.25) .. controls (264.58,239.75) and (262.55,241.78) .. (260.05,241.78) .. controls (257.55,241.78) and (255.52,239.75) .. (255.52,237.25) -- cycle ;
\draw  [fill={rgb, 255:red, 15; green, 8; blue, 8 }  ,fill opacity=1 ][line width=0.75]  (296.34,294.98) .. controls (296.34,292.48) and (298.37,290.45) .. (300.88,290.45) .. controls (303.38,290.45) and (305.41,292.48) .. (305.41,294.98) .. controls (305.41,297.49) and (303.38,299.52) .. (300.88,299.52) .. controls (298.37,299.52) and (296.34,297.49) .. (296.34,294.98) -- cycle ;
\draw  [fill={rgb, 255:red, 15; green, 8; blue, 8 }  ,fill opacity=1 ][line width=0.75]  (297.97,95.05) .. controls (297.97,92.55) and (300,90.52) .. (302.5,90.52) .. controls (305,90.52) and (307.03,92.55) .. (307.03,95.05) .. controls (307.03,97.55) and (305,99.58) .. (302.5,99.58) .. controls (300,99.58) and (297.97,97.55) .. (297.97,95.05) -- cycle ;

\draw (481.33,160.33) node [anchor=north west][inner sep=0.75pt]    {$\boldsymbol{N}$};
\draw (312.14,208.36) node [anchor=north west][inner sep=0.75pt]  [color={rgb, 255:red, 251; green, 244; blue, 244 }  ,opacity=1 ]  {$\textcolor[rgb]{0.07,0.01,0.01}{e}$};
\draw (372.6,208.4) node [anchor=north west][inner sep=0.75pt]    {${\textstyle f}$};
\draw (332.6,108.3) node [anchor=north west][inner sep=0.75pt]    {$a$};
\draw (160,383) node [anchor=north west][inner sep=0.75pt]    {$X_{1}$};
\draw (20,272.33) node [anchor=north west][inner sep=0.75pt]    {$X_{3}$};
\draw (1.67,387) node [anchor=north west][inner sep=0.75pt]    {$X_{2}$};

\end{tikzpicture}
    \caption{A typical three-dimensional finite volume cell where $e$ represents the cell center, $f$  the face center, $a$ a cell node, and $\boldsymbol{N}$ the outward unit normal vector.}
    \label{fig:finite volume cell }
\end{figure}

%% file: sections/numerical_flux_evaluation.tex
\subsection{Numerical Flux Evaluation}
The finite volume discretization presented in Eq.~\eqref{eq:semi-discretized} is based on a piecewise representation of the solution vector $\boldsymbol{\mathcal{U}}$ within each control volume \cite{hirsch1990numerical,toro2013riemann}. This discretization framework naturally leads to the formulation of a Riemann problem at every cell interface. To evaluate the numerical fluxes across these interfaces, the corresponding Riemann problems are solved using a newly developed Roe-type approximate Riemann solver, specifically designed and implemented for the present study.

Since Riemann problems are fundamentally one-dimensional, it is essential to transform the original system Eq.~\eqref{eq:first-order system} into an augmented one-dimensional system of equations at each cell interface. This transformation is facilitated by the rotational invariance property of the Eq.~\eqref{eq:first-order system}, which not only permits such a dimensional reduction in a finite volume framework but also supports the demonstration of the system’s hyperbolicity in time for the multi-dimensional case \cite{toro2013riemann}.

\subsubsection{Rotational Invariance}\label{subsec: rot_invariance}
Proposition:\ Consider a control volume with outward unit normal vector $\boldsymbol{N} = (N_1, N_2, N_3)$. The time-dependent, three-dimensional first-order system of equations (Eq.~\eqref{eq:first-order system}) is rotationally invariant, i.e., it satisfies
	\begin{equation}
		\underbrace{\boldsymbol{\mathcal{F}}_1 \, N_1+   \boldsymbol{\mathcal{F}}_2 \, N_2+ \boldsymbol{\mathcal{F}}_3 \, N_3
        }_{\displaystyle\boldsymbol{{{{\cal F}}}_N}}
        = \boldsymbol{T}^{-1} \boldsymbol{\mathcal{F}}_1(\boldsymbol{{T}}\boldsymbol{\mathcal{U}}) =: \boldsymbol{T}^{-1}\boldsymbol{\hat{\mathcal{F}_1}}.
		\label{eq:rotation invariance}
	\end{equation}
The rotation matrix $\boldsymbol{T}$ is defined as
	 \begin{equation}
		\boldsymbol{T} 
		\left.=\left(\begin{array}{ccc}{N_1}&{N_2}&{N_3}\\{t_{11}}&{t_{12}}&{t_{13}}\\{t_{21}}&{t_{22}}&{t_{23}}\\\end{array}\right.\right),
		\end{equation}
 with  $\boldsymbol{t_1}=(t_{11}, t_{12}, t_{13})$ and $\boldsymbol{t_2}=(t_{21}, t_{22}, t_{23})$ as unit orthogonal vectors tangent to the face.

\textit{Proof:}\\ It may be noted that the left-hand side (LHS) of Eq.~\eqref{eq:rotation invariance} can be written as,
\begin{equation}
	\boldsymbol{\mathcal{F}}_1 \, N_1+   \boldsymbol{\mathcal{F}}_2 \, N_2+ \boldsymbol{\mathcal{F}}_3 \, N_3 =
	\begin{bmatrix}
		-\boldsymbol{{P}} \boldsymbol{N}\\
		-\frac{1}{\rho_0} \boldsymbol{{p}} \otimes \boldsymbol{N}
	\end{bmatrix}.
\end{equation}
Now, we look at the expression of the right-hand side (RHS). Starting with 
	\begin{equation}
	\boldsymbol{\hat{\mathcal{F}_1}}= \boldsymbol{\mathcal{F}}_1 (\boldsymbol{\hat{\mathcal{U}}})=
	\begin{bmatrix}
		-\boldsymbol{\hat{P}} \boldsymbol{E}_1\\
		-\frac{1}{\rho_0} \boldsymbol{\hat{p}} \otimes \boldsymbol{E}_1
	\end{bmatrix}, 
    \label{eq : RHS}
	\end{equation}
where the rotated variables are defined as
\begin{equation}
	\boldsymbol{\hat{\mathcal{U}}} = 
	\begin{bmatrix}
		\boldsymbol{\hat{p}}\\
		\boldsymbol{\hat{F}}
	\end{bmatrix}
	=
	\begin{bmatrix}
		\boldsymbol{T} \boldsymbol{p}\\
		\boldsymbol{T} \boldsymbol{F} \boldsymbol{T}^{-1}
	\end{bmatrix},\quad \boldsymbol{\hat{p}} = \boldsymbol{Tp}, \quad \boldsymbol{\hat{P}}=\boldsymbol{TPT^{-1}}.
	\label{Eq: rotated_conservative_vector}
\end{equation}
Applying $\boldsymbol{T^{-1}}$ to the expression in Eq.~\eqref{eq : RHS} yields
\begin{equation}
	\boldsymbol{T}^{-1} \boldsymbol{\hat{\mathcal{F}_1}} = 
	\begin{bmatrix}
		- \boldsymbol{T}^{-1} (\boldsymbol{\hat{P}} \boldsymbol{E}_1)\\
		-\frac{1}{\rho_0} \boldsymbol{T}^{-1}(\boldsymbol{\hat{p}} \otimes \boldsymbol{E}_1) \boldsymbol{T}
	\end{bmatrix}=
	\begin{bmatrix}
		-\boldsymbol{{P}} \boldsymbol{N}\\
		-\frac{1}{\rho_0} \boldsymbol{{p}} \otimes \boldsymbol{N}
	\end{bmatrix} = LHS.
\end{equation}
This result confirms that the system is rotationally invariant.


\subsubsection{ Hyperbolicity in Time}
\input{figures/rotational_invariant}

Consider a locally rotated coordinate system 
 $(\hat{X}_1,\hat{X}_2,\hat{X}_3)$, as schematically illustrated in Figure~\ref{fig:tikz_example}, where the $\hat{X}_1$-axis is aligned with the interface normal, and the $\hat{X}_2$- and $\hat{X}_3$-axes lie tangential to the interface. In this rotated frame, only the flux component normal to the interface, $\hat{\boldsymbol{\mathcal{F}_1}}$, directed along the $\hat{X}_1$ axis, contributes to the finite volume flux computation, while the other flux components do not. Consequently, the flux evaluation using a Riemann solver requires only the following reduced form of governing equations, known as the augmented one-dimensional system \cite{toro2013riemann}: 
  \begin{equation}
 	\frac{\partial \hat{\boldsymbol{\mathcal{U}}}}{\partial t} + \frac{\partial\hat{\boldsymbol{\mathcal{F}_1}}}{\partial\hat{{X}_1}} = 0.
 	\label{eq:augmented 1d system}
 \end{equation}
Here, the conservative vector $\hat{\boldsymbol{\mathcal{U}}}$ and flux vector $\hat{\boldsymbol{\mathcal{F}_1}}$ are defined in Eq.~\eqref{Eq: rotated_conservative_vector} and Eq.~\eqref{eq : RHS} respectively.

We can write this system in a quasi-linear form:
\begin{equation}
	\frac{\partial \boldsymbol{\hat{\mathcal{U}}}}{\partial t} + \boldsymbol{\hat{\cal{A}}}\frac{\partial\boldsymbol{\hat{\mathcal{U}}}}{\partial{\hat{X}_1}} = 0
\end{equation}
where, Jacobian of the flux vector  $\boldsymbol{\hat{\cal{A}}}$ is given as:
\begin{equation}
	\boldsymbol{\mathcal{\hat{A}}} = \frac{\partial \boldsymbol{\hat{\cal{F}}_1}}{\partial\boldsymbol{\hat{\mathcal{U}}}} =\left(\begin{array}{cc}-\frac{\partial(\boldsymbol{\hat{P}}\boldsymbol{E}_1)}{\partial \boldsymbol{\hat{p}}}&-\frac{\partial(\boldsymbol{\hat{P}}\boldsymbol{E}_1)}{\partial \boldsymbol{\hat{F}}}\\
		-\frac{\partial\left(\frac{1}{\rho_0}\boldsymbol{\hat{p}}\otimes\boldsymbol{E}_1\right)}{\partial \boldsymbol{\hat{p}}}&-\frac{\partial\left(\frac{1}{\rho_0}\boldsymbol{\hat{p}}\otimes\boldsymbol{E}_1\right)}{\partial \boldsymbol{\hat{F}}}\end{array}\right)=
	\left(\begin{array}{cc}0_{3\times3}&-\boldsymbol{\mathcal{\hat{C}}}_E\\-\frac1{\rho_0}\boldsymbol{I}_E&\mathbf{0}_{9\times9}\end{array}\right),
	\label{eq: jacobian}
\end{equation}
where,  
\begin{equation*}
	[\boldsymbol{\mathcal{C}}_E]_{ijJ}=\frac{\partial \hat{P}_{iI}}{\partial \hat{F}_{jJ}}[\boldsymbol{E}_1]_I,\quad[\boldsymbol{I}_E]_{iIk}=\delta_{ik}[\boldsymbol{E}_1]_I.
\end{equation*}
Note that $\boldsymbol{\mathcal{\hat{C}}}_E$ denotes the $\boldsymbol{E}_1$ component of the constitutive material fourth order tensor $\boldsymbol{\mathcal{\hat{C}}}=\frac{\partial \boldsymbol{\hat{P}}}{\partial \boldsymbol{\hat{F}}}$, where we have dropped the subscript from $E$  for simplicity. This convention will be carried out for the rest of this section.

Considering the linear elastic materials, the fourth-order constitutive tensor yields 
\begin{equation}
\boldsymbol{\mathcal{\hat{C}}}=\lambda\mathbf{I}\otimes\mathbf{I}+\mu(\mathcal{I}+\overline{\mathcal{I}}),
\end{equation}
where,
\begin{equation}
\boldsymbol{I}\otimes\boldsymbol{I}=\delta_{iI}\delta_{jJ};\quad{\mathcal{I}}=\delta_{ij}\delta_{IJ};\quad\overline{\mathcal{I}}=\delta_{iJ}\delta_{Ij}.
\end{equation}
 The right eigen-decomposition equation is given by
\begin{equation}
    \boldsymbol{\mathcal{\hat{A}}} \boldsymbol{\mathcal{\hat{K}}}_{\alpha} =U_{\alpha}   \boldsymbol{\mathcal{\hat{K}}}_{\alpha}.
    \label{eq: right egeinvector equation}
\end{equation} 
The eigenvalues of this system are three pairs of
\begin{equation}
\begin{aligned}
        U_{1,2} &= \pm U_p,\\
        U_{3,4} = U_{5,6} &= \pm U_s,
\end{aligned}
\label{eq: eigenvalues}
\end{equation}
where
\begin{equation}
    U_p = \sqrt{\frac{\lambda+2\mu}{\rho_0}}, \quad U_s = \sqrt{\frac{\mu}{\rho_0}}.
\end{equation}
The rest of the eigenvalues are equal to zero. Note that the values of $U_P$ and $U_S$  are always positive. 
The corresponding right eigenvectors are found to be
\begin{equation}
\left(
\begin{array}{cccccc}
 1 & 1 & 0 & 0 & 0 & 0 \\
 0 & 0 & 1 & 0 & 1 & 0 \\
 0 & 0 & 0 & 1 & 0 & 1 \\
 -\frac{1}{\rho  \text{Up}} & \frac{1}{\rho  \text{Up}} & 0 & 0 & 0 & 0 \\
 0 & 0 & 0 & 0 & 0 & 0 \\
 0 & 0 & 0 & 0 & 0 & 0 \\
 0 & 0 & -\frac{1}{\rho  \text{Us}} & 0 & \frac{1}{\rho  \text{Us}} & 0 \\
 0 & 0 & 0 & 0 & 0 & 0 \\
 0 & 0 & 0 & 0 & 0 & 0 \\
 0 & 0 & 0 & -\frac{1}{\rho  \text{Us}} & 0 & \frac{1}{\rho  \text{Us}} \\
 0 & 0 & 0 & 0 & 0 & 0 \\
 0 & 0 & 0 & 0 & 0 & 0 \\
\end{array}
\right).
\end{equation}
Finally, to find the rest of the eigenvectors that correspond to the zero eigenvalues, we solve the null space for the whole system to produce the following set of eigenvectors
\begin{equation}
    \left(
\begin{array}{cccccc}
 0 & 0 & 0 & 0 & 0 & 0 \\
 0 & 0 & 0 & 0 & 0 & 0 \\
 0 & 0 & 0 & 0 & 0 & 0 \\
 -\frac{\lambda }{\lambda +2 \mu } & 0 & 0 & 0 & -\frac{\lambda }{\lambda +2 \mu } & 0 \\
 0 & 0 & 0 & 0 & 0 & -1 \\
 0 & 0 & -1 & 0 & 0 & 0 \\
 0 & 0 & 0 & 0 & 0 & 1 \\
 0 & 0 & 0 & 0 & 1 & 0 \\
 0 & 0 & 0 & 1 & 0 & 0 \\
 0 & 0 & 1 & 0 & 0 & 0 \\
 0 & 1 & 0 & 0 & 0 & 0 \\
 1 & 0 & 0 & 0 & 0 & 0 \\
\end{array}
\right).
\end{equation}

Thus, the hyperbolicity in time of the three-dimensional system (Eq.~\eqref{eq:first-order system}) can be established \cite{toro2013riemann}, since the augmented one-dimensional system (Eq.~\eqref{eq:augmented 1d system}) possesses twelve real eigenvalues and a complete set of twelve linearly independent eigenvectors, and the system satisfies the rotational invariance property demonstrated in Section~\ref{subsec: rot_invariance}.


\subsubsection{Solution Reconstruction}
In the cell-centered finite volume method, the solution vector $\boldsymbol{\mathcal{U}}$, representing the volume-averaged conserved variables, is defined at the centroid of each control volume. However, the numerical fluxes required for updating the solution are evaluated at the cell faces, where the values of the conserved variables are not readily available. This necessitates interpolating either the fluxes or the flow variables to the cell faces—a process commonly referred to as solution reconstruction.

A piecewise constant reconstruction assumes a uniform solution within each control volume, leading to a first-order accurate scheme. In contrast, linear reconstruction provides second-order accuracy by accounting for solution gradients. Nevertheless, higher-order reconstructions can produce non-physical oscillations near discontinuities or introduce spurious extrema in smooth regions of the solution. To mitigate these issues, the Monotonic Upstream-centered Schemes for Conservation Laws (MUSCL) approach \cite{toro2013riemann,hirsch1990numerical, leveque2002finite} is employed, which incorporates slope limiters to maintain stability and avoid unphysical overshoots.

In the present formulation, the reconstruction begins with a first-order Taylor series expansion of the solution about the cell centroid $\boldsymbol{X}_e$ to approximate the variable ${\cal{\hat{U}}}_e(\boldsymbol{X})$ at any point $\boldsymbol{X}$ within the cell:
\begin{equation}
    {\cal{\hat{U}}}_e(\boldsymbol{X}) = {\cal{\hat{U}}}_e + \nabla{\cal{\hat{U}}}_e (\boldsymbol{X}-\boldsymbol{X}_e).
\end{equation}
Here, $\nabla{\cal{\hat{U}}}_e$ is the gradient of the solution, estimated using a least-squares minimization approach \cite{lee2013development}:
\begin{equation}
\nabla{\cal{\hat{U}}}_e=\left[\sum_{f=1}^{m}\boldsymbol{\nu}_{ef}\otimes\boldsymbol{\nu}_{ef}\right]^{-1}\sum_{f=1}^{m}\left(\frac{\mathcal{\hat{U}}_{f}-\mathcal{\hat{U}}_{e}}{d_{ef}}\right)\boldsymbol{\nu}_{ef};
\quad\boldsymbol{\nu}_{ef}=\frac{\boldsymbol{X}_{f}-\boldsymbol{X}_{e}}{d_{ef}}.
\end{equation}
The Venkatakrishnan limiter \cite{venkatakrishnan1995convergence}  is applied to the computed gradients to limit non-physical oscillations. The modified reconstruction with limiting takes the form:
\begin{equation}
    {\cal{\hat{U}}}_e(\boldsymbol{X}) = {\cal{\hat{U}}}_e + 
 \Phi_e \nabla{\cal{\hat{U}}}_e (\boldsymbol{X}-\boldsymbol{X}_e),
\end{equation}
where $\Phi_e$ is the limiter function defined as:
\begin{equation}
    \Phi_e = \min(\Phi_{ef}), \quad \forall f = 1,2,\ldots,M
\end{equation}
For each face, the face-specific limiter $\Phi_{ef}$ is computed as:
\begin{equation}\left.\Phi_{ef}=
\begin{cases}
\min\left(1,\phi\left(\frac{\mathcal{\hat{U}}^{\max}-\mathcal{\hat{U}}_{e}}{\mathcal{\hat{U}}_{ef}-\mathcal{\hat{U}}_{e}}\right)\right),&\mathrm{if}\quad\mathcal{\hat{U}}_{ef}-\mathcal{\hat{U}}_{e}>0\\
\min\left(1,\phi\left(\frac{\mathcal{\hat{U}}^{\min}-\mathcal{\hat{U}}_{e}}{\mathcal{\hat{U}}_{ef}-\mathcal{\hat{U}}_{e}}\right)\right),&\mathrm{if}\quad\mathcal{\hat{U}}_{ef}-\mathcal{\hat{U}}_{e}<0\\
1,&\mathrm{if}\quad\mathcal{\hat{U}}_{ef}-\mathcal{\hat{U}}_{e}=0
\end{cases}\right.\end{equation}
where
\begin{equation}\mathcal{\hat{U}}^{\mathrm{min}}=\min(\mathcal{\hat{U}}_{e},\mathcal{\hat{U}}_{f})\quad\mathrm{and}\quad\mathcal{\hat{U}}^{\mathrm{max}}=\max(\mathcal{\hat{U}}_{e},\mathcal{\hat{U}}_{f}), \quad  \forall f=1,\ldots,m,\end{equation}
and the function $\phi\left(\eta\right)$ is given by:
\begin{equation}\phi\left(\eta\right)=\frac{\eta^{2}+2\eta}{\eta^{2}+\eta+2} .\end{equation}
This reconstruction procedure provides limited, high-resolution gradients that preserve accuracy in smooth regions while avoiding spurious oscillations near sharp gradients or discontinuities.



\subsubsection{Flux Evaluation using Riemann Solver}
\label{subsubsec:Riemann Solver}
In the finite volume method, the reconstruction of conservative variables within each control volume inherently introduces discontinuities at cell interfaces. This results in distinct left and right states, denoted by  $\hat{\boldsymbol{\mathcal{U}}}_L$ and $\hat{\boldsymbol{\mathcal{U}}}_R$, respectively, on either side of each interface. Consequently, the method naturally gives rise to a local Riemann problem at each cell face. The solution of this Riemann problem facilitates the evaluation of the numerical flux, $\hat{\boldsymbol{\cal F}}_1(\hat{\boldsymbol{\cal U}}) = \hat{\boldsymbol{\cal F}}_1(\hat{\boldsymbol{\cal U}}_L,\hat{\boldsymbol{\cal U}}_R)$, across the interface. A Riemann solver is typically employed to compute this flux.

In the present work, we propose a Roe-type Riemann solver for the augmented one-dimensional equations, Eq.~\eqref{eq:augmented 1d system}, to compute interface flux in the rotated coordinate system as:
\begin{equation}
	\boldsymbol {\hat{\mathcal{F}_1}} = \frac{1}{2}\left[\boldsymbol {\hat{\mathcal{F}_1}}(\boldsymbol{\hat{\mathcal{U}}}_L) +\boldsymbol {\hat{\mathcal{F}_1}}
	(\boldsymbol{\hat{\mathcal{U}}}_R) \right]
	-
	\frac{1}{2}
	\sum_{j} \Tilde{\alpha_j} |\Tilde{U_j}|\Tilde{\boldsymbol{\cal{K}}}^j .
\end{equation}
Here, $\Tilde{U_j}$ and $\Tilde{\boldsymbol{\cal{K}}}^j$ represent the $j^{th}$ eigenvalue and right eigenvector of the Roe-averaged flux Jacobian matrix $\boldsymbol{\hat{\mathcal{A}}}$. In the context of the linear elasticity model used here, these quantities are constants determined solely by solid material properties; hence, no averaging is necessary. $\Tilde{\alpha}_j$ denotes the wave strength of the $j^{th}$ wave.  The wave strength $\tilde{\alpha}_j$ corresponding to the j-th wave is obtained by projecting the jump in the conservative variable vector, $\Delta \boldsymbol{\mathcal{\hat{U}} = \boldsymbol{\mathcal{\hat{U}}}_R - \boldsymbol{\mathcal{\hat{U}}}}_L$, onto the corresponding right eigenvectors, as follows \cite{toro2013riemann}:
\begin{equation}
	\Delta \boldsymbol{\mathcal{\hat{U}}} = \sum_{j=1}^{12} \tilde{\alpha}_j \boldsymbol{\tilde{\mathcal{K}}}^j.
	\label{Eq:wave_strength}
\end{equation}
Solving Eq.~\eqref{Eq:wave_strength} yields the expressions for the wave strengths $\tilde{\alpha}_j$:
 \begin{equation}
	\begin{aligned}
		\tilde{\alpha}_1 &= \frac{1}{2} \left(-\rho\,U_P\, \Delta \hat{F}_{11} - \frac{\lambda}{U_P}(\Delta \hat{F}_{22}+\Delta \hat{F}_{33}) +\Delta \hat{p}_1 \right),\\
		\tilde{\alpha}_2 &= \frac{1}{2} \left( \rho\,U_P\, \Delta \hat{F}_{11} + \frac{\lambda}{U_P}(\Delta \hat{F}_{22}+\Delta \hat{F}_{33}) +\Delta \hat{p}_1 \right),\\
		\tilde{\alpha}_3 &= \frac{1}{2} \left( -\rho\, U_S(\Delta\hat{F}_{12} + \Delta\hat{F}_{21} )+\Delta\hat{p}_2\right),\\
		\tilde{\alpha}_4 &= \frac{1}{2} \left( -\rho\, U_S(\Delta\hat{F}_{13} + \Delta\hat{F}_{31} )+\Delta\hat{p}_3\right),\\
		\tilde{\alpha}_5 &= \frac{1}{2} \left( \rho\, U_S(\Delta\hat{F}_{12} + \Delta\hat{F}_{21} )+\Delta\hat{p}_2\right),\\
		\tilde{\alpha}_6 &= \frac{1}{2} \left( \rho\, U_S(\Delta\hat{F}_{13} + \Delta\hat{F}_{31} )+\Delta\hat{p}_3\right).
	\end{aligned}
\end{equation}
Here, $\Delta(.)$ denotes the difference in the respective quantities between the right and left states.

Once the flux $\boldsymbol{\mathcal{\hat{F}}}_1$ is computed in each interface in the rotated coordinate system, it is transformed back to the original coordinate system using the inverse transformation. These transformed fluxes are then substituted into the finite volume formulation, Eq.(\ref{eq:semi-discretized}),  resulting in the semi-discrete evolution equation:
\begin{equation}
	\frac{\mathrm{d} \boldsymbol{\cal{U}}_e}{\mathrm{d} t}  = \underbrace{ -\frac{1}{\Omega_e }\sum_{f=1}^M (\boldsymbol{T}^{-1}\boldsymbol{\hat{\mathcal{F}_1}} \, \Delta S)_f+  \boldsymbol{\cal{S}}_e }_{\displaystyle\boldsymbol{ R}\left(\boldsymbol{{{{\cal U}}}}\right)}.
	\label{eq:semi-discretized-final}
\end{equation}

%% file: figures/rotational_invariant.tex
\begin{figure}[ht]
    \centering
\tikzset{every picture/.style={line width=1pt}} 

\begin{tikzpicture}[x=0.5pt,y=0.5pt,yscale=-1,xscale=1,every node/.style={font=\footnotesize}]

\draw  [color={rgb, 255:red, 0; green, 0; blue, 0 }  ,draw opacity=0 ] (152.12,183.64) -- (187.91,104.44) -- (322.07,53.8) -- (376.9,199.05) -- (341.1,278.25) -- (206.95,328.89) -- cycle ; \draw  [color={rgb, 255:red, 0; green, 0; blue, 0 }  ,draw opacity=0 ] (322.07,53.8) -- (286.28,133) -- (152.12,183.64) ; \draw  [color={rgb, 255:red, 0; green, 0; blue, 0 }  ,draw opacity=0 ] (286.28,133) -- (341.1,278.25) ;
\draw [color={rgb, 255:red, 0; green, 0; blue, 0 }  ,draw opacity=0.41 ] [dash pattern={on 4.5pt off 4.5pt}]  (241.73,248.6) -- (375.75,196.03) ;
\draw [color={rgb, 255:red, 0; green, 0; blue, 0 }  ,draw opacity=0.41 ] [dash pattern={on 4.5pt off 4.5pt}]  (241.73,248.6) -- (206.95,328.89) ;
\draw [color={rgb, 255:red, 0; green, 0; blue, 0 }  ,draw opacity=0.41 ] [dash pattern={on 4.5pt off 4.5pt}]  (187.91,104.44) -- (241.73,248.6) ;
\draw  [fill={rgb, 255:red, 15; green, 8; blue, 8 }  ,fill opacity=1 ][line width=0.75]  (261.44,192.51) .. controls (260.77,190.72) and (261.59,188.76) .. (263.29,188.12) .. controls (264.98,187.48) and (266.9,188.4) .. (267.58,190.19) .. controls (268.25,191.97) and (267.42,193.94) .. (265.73,194.58) .. controls (264.03,195.22) and (262.11,194.29) .. (261.44,192.51) -- cycle ;
\draw  [line width=3.75]  (330.1,166.59) .. controls (329.77,165.72) and (330.17,164.77) .. (331,164.46) .. controls (331.82,164.15) and (332.75,164.6) .. (333.08,165.46) .. controls (333.4,166.33) and (333,167.29) .. (332.18,167.6) .. controls (331.36,167.91) and (330.42,167.46) .. (330.1,166.59) -- cycle ;
\draw  [line width=3.75]  (285.93,136.59) .. controls (285.6,135.73) and (286,134.77) .. (286.83,134.46) .. controls (287.65,134.15) and (288.58,134.6) .. (288.91,135.47) .. controls (289.24,136.33) and (288.83,137.29) .. (288.01,137.6) .. controls (287.19,137.91) and (286.26,137.46) .. (285.93,136.59) -- cycle ;
\draw  [line width=3.75]  (196.01,218.2) .. controls (195.68,217.33) and (196.08,216.38) .. (196.9,216.07) .. controls (197.73,215.75) and (198.66,216.2) .. (198.99,217.07) .. controls (199.31,217.94) and (198.91,218.89) .. (198.09,219.2) .. controls (197.27,219.51) and (196.33,219.06) .. (196.01,218.2) -- cycle ;
\draw  [dash pattern={on 4.5pt off 4.5pt}]  (197.5,217.63) -- (267.16,190.82) -- (267.48,190.7) -- (331.59,166.03) ;
\draw  [dash pattern={on 4.5pt off 4.5pt}]  (237.67,120.24) -- (290.37,263.18) ;
\draw  [line width=3.75]  (236.18,120.8) .. controls (235.85,119.93) and (236.25,118.98) .. (237.07,118.67) .. controls (237.9,118.36) and (238.83,118.81) .. (239.16,119.67) .. controls (239.48,120.54) and (239.08,121.49) .. (238.26,121.8) .. controls (237.43,122.11) and (236.5,121.66) .. (236.18,120.8) -- cycle ;
\draw  [line width=2.25]  (288.88,263.75) .. controls (288.55,262.88) and (288.95,261.93) .. (289.78,261.62) .. controls (290.6,261.31) and (291.53,261.76) .. (291.86,262.62) .. controls (292.19,263.49) and (291.78,264.44) .. (290.96,264.75) .. controls (290.14,265.06) and (289.21,264.61) .. (288.88,263.75) -- cycle ;
\draw  [line width=3.75]  (280.34,150.8) .. controls (280.02,149.93) and (280.42,148.98) .. (281.24,148.67) .. controls (282.06,148.35) and (283,148.8) .. (283.32,149.67) .. controls (283.65,150.54) and (283.25,151.49) .. (282.43,151.8) .. controls (281.6,152.11) and (280.67,151.66) .. (280.34,150.8) -- cycle ;
\draw  [line width=3.75]  (245.69,233.02) .. controls (245.37,232.16) and (245.77,231.2) .. (246.59,230.89) .. controls (247.41,230.58) and (248.35,231.03) .. (248.67,231.9) .. controls (249,232.77) and (248.6,233.72) .. (247.78,234.03) .. controls (246.95,234.34) and (246.02,233.89) .. (245.69,233.02) -- cycle ;
\draw    (247.18,232.46) -- (226.13,280.15) ;
\draw [shift={(225.32,281.98)}, rotate = 293.82] [color={rgb, 255:red, 0; green, 0; blue, 0 }  ][line width=0.75]    (10.93,-3.29) .. controls (6.95,-1.4) and (3.31,-0.3) .. (0,0) .. controls (3.31,0.3) and (6.95,1.4) .. (10.93,3.29)   ;
\draw  [dash pattern={on 4.5pt off 4.5pt}]  (281.83,150.23) -- (247.18,232.46) ;
\draw  [fill={rgb, 255:red, 15; green, 8; blue, 8 }  ,fill opacity=1 ][line width=0.75]  (328.52,167.19) .. controls (327.84,165.4) and (328.67,163.44) .. (330.37,162.8) .. controls (332.06,162.16) and (333.98,163.08) .. (334.66,164.87) .. controls (335.33,166.65) and (334.5,168.62) .. (332.81,169.26) .. controls (331.11,169.9) and (329.19,168.97) .. (328.52,167.19) -- cycle ;
\draw  [fill={rgb, 255:red, 15; green, 8; blue, 8 }  ,fill opacity=1 ][line width=0.75]  (278.76,151.39) .. controls (278.09,149.61) and (278.92,147.64) .. (280.61,147) .. controls (282.31,146.36) and (284.23,147.29) .. (284.9,149.07) .. controls (285.58,150.86) and (284.75,152.82) .. (283.05,153.46) .. controls (281.36,154.1) and (279.44,153.18) .. (278.76,151.39) -- cycle ;
\draw  [fill={rgb, 255:red, 15; green, 8; blue, 8 }  ,fill opacity=1 ][line width=0.75]  (284.35,137.19) .. controls (283.68,135.4) and (284.5,133.44) .. (286.2,132.8) .. controls (287.9,132.16) and (289.82,133.09) .. (290.49,134.87) .. controls (291.16,136.66) and (290.33,138.62) .. (288.64,139.26) .. controls (286.94,139.9) and (285.02,138.97) .. (284.35,137.19) -- cycle ;
\draw  [fill={rgb, 255:red, 15; green, 8; blue, 8 }  ,fill opacity=1 ][line width=0.75]  (194.43,218.79) .. controls (193.75,217.01) and (194.58,215.04) .. (196.28,214.4) .. controls (197.97,213.76) and (199.89,214.69) .. (200.57,216.47) .. controls (201.24,218.26) and (200.41,220.22) .. (198.72,220.86) .. controls (197.02,221.5) and (195.1,220.58) .. (194.43,218.79) -- cycle ;
\draw  [fill={rgb, 255:red, 15; green, 8; blue, 8 }  ,fill opacity=1 ][line width=0.75]  (244.11,233.62) .. controls (243.44,231.84) and (244.27,229.87) .. (245.96,229.23) .. controls (247.66,228.59) and (249.58,229.52) .. (250.25,231.3) .. controls (250.93,233.09) and (250.1,235.05) .. (248.4,235.69) .. controls (246.71,236.33) and (244.79,235.41) .. (244.11,233.62) -- cycle ;
\draw  [fill={rgb, 255:red, 15; green, 8; blue, 8 }  ,fill opacity=1 ][line width=0.75]  (287.3,264.34) .. controls (286.63,262.56) and (287.45,260.59) .. (289.15,259.95) .. controls (290.84,259.31) and (292.76,260.24) .. (293.44,262.03) .. controls (294.11,263.81) and (293.28,265.78) .. (291.59,266.42) .. controls (289.89,267.06) and (287.97,266.13) .. (287.3,264.34) -- cycle ;
\draw  [fill={rgb, 255:red, 15; green, 8; blue, 8 }  ,fill opacity=1 ][line width=0.75]  (234.6,121.39) .. controls (233.92,119.61) and (234.75,117.64) .. (236.45,117) .. controls (238.14,116.36) and (240.06,117.29) .. (240.73,119.08) .. controls (241.41,120.86) and (240.58,122.83) .. (238.89,123.47) .. controls (237.19,124.11) and (235.27,123.18) .. (234.6,121.39) -- cycle ;
\draw  [color={rgb, 255:red, 0; green, 0; blue, 0 }  ,draw opacity=0.07 ][fill={rgb, 255:red, 150; green, 148; blue, 148 }  ,fill opacity=1 ] (152.2,183.86) -- (286.36,133.22) -- (341.33,278.85) -- (207.17,329.49) -- cycle ;
\draw  [color={rgb, 255:red, 0; green, 0; blue, 0 }  ,draw opacity=0.07 ][fill={rgb, 255:red, 234; green, 227; blue, 227 }  ,fill opacity=1 ] (185.65,105.3) -- (322.41,53.67) -- (288.88,132.02) -- (152.12,183.64) -- cycle ;
\draw  [color={rgb, 255:red, 0; green, 0; blue, 0 }  ,draw opacity=0.07 ][fill={rgb, 255:red, 211; green, 195; blue, 195 }  ,fill opacity=1 ] (340.94,277.82) -- (286.55,133.18) -- (322.07,53.8) -- (376.47,198.45) -- cycle ;
\draw    (331.59,166.03) -- (421.41,131.91) ;
\draw [shift={(423.28,131.2)}, rotate = 159.2] [color={rgb, 255:red, 0; green, 0; blue, 0 }  ][line width=0.75]    (10.93,-3.29) .. controls (6.95,-1.4) and (3.31,-0.3) .. (0,0) .. controls (3.31,0.3) and (6.95,1.4) .. (10.93,3.29)   ;
\draw    (331.51,165.81) -- (291.71,60.47) ;
\draw [shift={(291,58.6)}, rotate = 69.3] [color={rgb, 255:red, 0; green, 0; blue, 0 }  ][line width=0.75]    (10.93,-3.29) .. controls (6.95,-1.4) and (3.31,-0.3) .. (0,0) .. controls (3.31,0.3) and (6.95,1.4) .. (10.93,3.29)   ;
\draw    (331.51,165.81) -- (300.92,234.36) ;
\draw [shift={(300.11,236.19)}, rotate = 294.05] [color={rgb, 255:red, 0; green, 0; blue, 0 }  ][line width=0.75]    (10.93,-3.29) .. controls (6.95,-1.4) and (3.31,-0.3) .. (0,0) .. controls (3.31,0.3) and (6.95,1.4) .. (10.93,3.29)   ;
\draw    (331.51,165.81) -- (371.15,150.82) ;
\draw [shift={(373.02,150.11)}, rotate = 159.28] [color={rgb, 255:red, 0; green, 0; blue, 0 }  ][line width=0.75]    (10.93,-3.29) .. controls (6.95,-1.4) and (3.31,-0.3) .. (0,0) .. controls (3.31,0.3) and (6.95,1.4) .. (10.93,3.29)   ;
\draw    (331.59,166.03) -- (316.94,197.93) ;
\draw [shift={(316.11,199.75)}, rotate = 294.66] [color={rgb, 255:red, 0; green, 0; blue, 0 }  ][line width=0.75]    (10.93,-3.29) .. controls (6.95,-1.4) and (3.31,-0.3) .. (0,0) .. controls (3.31,0.3) and (6.95,1.4) .. (10.93,3.29)   ;
\draw    (331.59,166.03) -- (318.1,131.25) ;
\draw [shift={(317.38,129.38)}, rotate = 68.81] [color={rgb, 255:red, 0; green, 0; blue, 0 }  ][line width=0.75]    (10.93,-3.29) .. controls (6.95,-1.4) and (3.31,-0.3) .. (0,0) .. controls (3.31,0.3) and (6.95,1.4) .. (10.93,3.29)   ;
\draw  [color={rgb, 255:red, 0; green, 0; blue, 0 }  ,draw opacity=0 ][fill={rgb, 255:red, 206; green, 47; blue, 47 }  ,fill opacity=1 ] (333.88,170.23) .. controls (333.09,171.57) and (331.42,170.77) .. (330.15,168.45) .. controls (328.88,166.13) and (328.5,163.16) .. (329.29,161.83) .. controls (330.09,160.49) and (331.76,161.29) .. (333.02,163.61) .. controls (334.29,165.93) and (334.68,168.89) .. (333.88,170.23) -- cycle ;
\draw    (112.09,344.71) -- (111.15,189.61) ;
\draw [shift={(111.14,187.61)}, rotate = 89.65] [color={rgb, 255:red, 0; green, 0; blue, 0 }  ][line width=0.75]    (10.93,-3.29) .. controls (6.95,-1.4) and (3.31,-0.3) .. (0,0) .. controls (3.31,0.3) and (6.95,1.4) .. (10.93,3.29)   ;
\draw    (112.09,344.71) -- (269.18,345.18) ;
\draw [shift={(271.18,345.18)}, rotate = 180.17] [color={rgb, 255:red, 0; green, 0; blue, 0 }  ][line width=0.75]    (10.93,-3.29) .. controls (6.95,-1.4) and (3.31,-0.3) .. (0,0) .. controls (3.31,0.3) and (6.95,1.4) .. (10.93,3.29)   ;
\draw    (112.09,344.71) -- (34.86,427.24) ;
\draw [shift={(33.49,428.7)}, rotate = 313.1] [color={rgb, 255:red, 0; green, 0; blue, 0 }  ][line width=0.75]    (10.93,-3.29) .. controls (6.95,-1.4) and (3.31,-0.3) .. (0,0) .. controls (3.31,0.3) and (6.95,1.4) .. (10.93,3.29)   ;

\draw (394.19,112.77) node [anchor=north west][inner sep=0.75pt]  [rotate=-359.93]  {$\boldsymbol{N}$};
\draw (333.78,124.07) node [anchor=north west][inner sep=0.75pt]  [rotate=-359.93]  {$\hat{X}_{1}$};
\draw (266.05,58.4) node [anchor=north west][inner sep=0.75pt]  [rotate=-359.93]  {$t_{2}$};
\draw (277.77,211.57) node [anchor=north west][inner sep=0.75pt]  [rotate=-359.93]  {$t_{1}$};
\draw (286.59,128.95) node [anchor=north west][inner sep=0.75pt]  [rotate=-359.93]  {$\hat{X}_{3}$};
\draw (321.79,182.01) node [anchor=north west][inner sep=0.75pt]  [rotate=-359.93]  {$\hat{X}_{2}$};
\draw (273.28,333.02) node [anchor=north west][inner sep=0.75pt]    {$X_{1}$};
\draw (99.44,166.63) node [anchor=north west][inner sep=0.75pt]    {$X_{3}$};
\draw (12.39,422.32) node [anchor=north west][inner sep=0.75pt]    {$X_{2}$};

\end{tikzpicture}

    \caption{Locally-rotated coordinate system $\hat{\boldsymbol{X}}$}
    \label{fig:tikz_example}
\end{figure}

%% file: sections/time-integration.tex

\subsection{Integration of Time Derivative}
Time integration of Eq.~\eqref{eq:semi-discretized-final} is performed using the second-order Strong Stability Preserving Runge-Kutta (SSPRK) method, which offers improved accuracy and stability for hyperbolic systems \cite{gottlieb2001strong}. This two-stage explicit scheme is given by:
\begin{equation}
	\begin{array}{l}
		{{\boldsymbol{{{{\cal U}}}}^{(1)}=\boldsymbol{{{{\cal U}}}}^{(n)}+\Delta t\,\boldsymbol{ R}\left(\boldsymbol{{{{\cal U}}}}^{(n)}\right),}}\\ 
		{{\boldsymbol{{{{\cal U}}}}^{(2)}=\boldsymbol{{{{\cal U}}}}^{(1)}+\Delta t\,\boldsymbol{ R}\left(\boldsymbol{{{{\cal U}}}}^{(1)}\right),}}\\
		{{\boldsymbol{{{{\cal U}}}}^{n+1}=\frac12\left(\boldsymbol{{{{\cal U}}}}^{(n)}+\boldsymbol{{{{\cal U}}}}^{(2)}\right),}}\end{array} 
\end{equation}
where $\boldsymbol{ R}\left(\boldsymbol{{{{\cal U}}}}\right)$ denotes the residual vector (i.e. RHS of Eq.~\eqref{eq:semi-discretized-final}) evaluated using conservative variables at each stage. This formulation ensures second-order temporal accuracy while preserving strong stability properties under suitable time step restrictions.

The time step $\Delta t$ is constrained by the Courant-Friedrichs-Lewy (CFL) condition \cite{courant1928partial} to maintain numerical stability:
\begin{equation}
	\Delta t = \alpha_{CFL}\frac{h_{min}}{U_{p,max}},
\end{equation}
where \(h_{min}\) is the minimum local grid size, \(U_{p,max}\) is the maximum characteristic wave speed in the domain, and \(\alpha_{CFL}\) is a user-defined stability coefficient (typically \(\alpha_{CFL} < 1 \))

%% file: sections/boundary_conditions.tex
\subsection{Boundary Conditions}
\label{subsec:Boundary Conditions}
To evaluate the numerical boundary fluxes, we utilize the boundary conditions derived using the Contact algorithm derived by Lee et al.\cite{lee2013development}, which evaluates the boundary traction vector $\boldsymbol{t} = \boldsymbol{P} \boldsymbol{N}$ and linear momentum $\boldsymbol{p}$.  Calculating these quantities, we can substitute them back into the boundary flux vector:
\begin{equation}
    \boldsymbol{\mathcal{F}}_N =
    \begin{bmatrix}
        \boldsymbol{P} \boldsymbol{N}\\
        \frac{1}{\rho_0} \boldsymbol{p} \otimes \boldsymbol{N}
    \end{bmatrix},
\end{equation}
where the subscript $N$ refers to the normal flux vector (see Eq.~\eqref{eq:rotation invariance}). In general, there are two main types of boundary conditions: moving boundary and traction boundary, which are described below.
 \subsubsection{Moving boundary}
The boundary has a prescribed linear momentum $ \boldsymbol{p}_B$. Thus, the linear momentum and the traction are given by:
\begin{equation}
\begin{aligned}
        \boldsymbol{p}_f &= \boldsymbol{p}_B,\
        \boldsymbol{t}_f &= \boldsymbol{t}_L + \boldsymbol{S}_{\boldsymbol{p}}(\boldsymbol{p}_B - \boldsymbol{p}_L).  
\end{aligned}
\end{equation}
In a fixed boundary, the linear momentum at the boundary is zero ( $\boldsymbol{p}_B=0$). 
 \subsubsection{Traction boundary}
 In this case, the boundary has a prescribed traction $ \boldsymbol{t}_B$. Thus, the linear momentum and the traction are given by:
\begin{equation}
\begin{aligned}
        \boldsymbol{p}_f &= \boldsymbol{p}_L + \boldsymbol{S}_{\boldsymbol{t}}(\boldsymbol{t}_b - \boldsymbol{t}_L),  \\
        \boldsymbol{t}_f &= \boldsymbol{t}_B.
\end{aligned}
\end{equation}
In the case of the free surface, i.e., traction-free,  the traction is set to zero ( $\boldsymbol{t}_B=0$). 

%% file: sections/solution_algorithm.tex
Section~\ref{subsec: rot_invariance}\section{Solution Algorithm}
\label{sec:Solution Algorithm}
In addition to the finite volume method discretization steps involving linear reconstruction, flux computation, and time marching, solving the solid dynamic system demands additional considerations. One critical requirement is that the system must satisfy the compatibility condition, ensuring that the deformation gradient is curl-free  $\boldsymbol{\nabla} \times \boldsymbol{F}$. This condition can be enforced using various approaches, such as employing specific numerical schemes or introducing auxiliary constraints; however, in this work, we adopt the C-TOUCH algorithm developed by Haider et al. \cite{haider2017first}.
Secondly, the requirement for a curl-free algorithm, combined with using the C-TOUCH framework, means that the deformation gradient is no longer directly computed as the material gradient of the current geometry. As a result, the conservation of angular momentum is not inherently satisfied. To address this issue, an additional algorithm is necessary to ensure the conservation of angular momentum \cite{lee2013development,haider2017first}.
\subsection{Curl-Free Algorithm}
The curl-free deformation field can be ensured by making the time update of the deformation gradient dependent on the nodal linear momentum. This necessitates modifying the semi-discrete equation of the deformation gradient, Eq.~\eqref{eq : semi-discret deformation}, as follows:
\begin{equation}
    \frac{\mathrm{d} \boldsymbol{F}_e}{\mathrm{d} t} =  - \frac{1}{\Omega_e }  \sum_{f \in \Lambda_e^f} (\frac{1}{\rho_0}\boldsymbol{\bar{p}} \otimes\boldsymbol{N} \, \Delta S)_f,
\end{equation}
where $\boldsymbol{\bar{p}}$ represents the filtering linear momentum, defined as  
\begin{equation}  
    \boldsymbol{\bar{p}} = \frac{1}{\Lambda_f^a} \sum_{a \in \Lambda_f^a} \boldsymbol{p}_a ,
\end{equation}  
where $\boldsymbol{p}_a$ is the nodal linear momentum, and $\Lambda_f^a$ denotes the set of nodes of face $f$. The filtering linear momentum is calculated using the C-TOUCH algorithm outlined in \cite{haider2017first}. This process involves a sequence of interpolation steps between node, face center, and cell center fields within the discretized geometry, as detailed in the following steps:

\textit{step one, evaluate the face center linear momentum}: After the evaluation of the flux vector using the Riemann solver, it is possible to extract the linear momentum evaluated at the face center, noticing that we have already evaluated the quantity
\begin{equation}
    \boldsymbol{\mathcal{F}_1}_{\boldsymbol{F} } := (\frac{1}{\rho_0} \boldsymbol{p} \otimes \boldsymbol{N})_f,
\end{equation}
where ${\boldsymbol{\mathcal{F}_1}}_{\boldsymbol{F}}$ is the flux term of deformation gradient equation. Using simple algebraic manipulation, it can be shown that
\begin{equation}
    \boldsymbol{p}_f = \frac{\rho_0}{||\boldsymbol{N}||^2} {\boldsymbol{\mathcal{F}_1}}_{\boldsymbol{F}} \boldsymbol{N},
\end{equation}

\textit{Step two, obtain averaged linear momentum:} using simple interpolation, the averaged linear momentum at the cell center is evaluated by 
\begin{equation}
    \boldsymbol{\bar{p}}_e = \frac{1}{\Lambda_e^f} \sum _{ f \in {\Lambda_e^f}} \boldsymbol{p}_f,
\end{equation}

\textit{Step three,  compute elemental nodal linear momentum:} following the same concept of linear reconstruction and interpolating the cell center values to the face center, here the interpolation is done from  the cell center to the nodes
\begin{equation}
    \boldsymbol{p}_{ea} = \boldsymbol{\bar{p}}_e + \nabla \boldsymbol{\bar{p}}_e (\boldsymbol{X}_a -\boldsymbol{X}_e)
\end{equation}

\textit{Step four,  obtain continuous nodal linear momentum:} This is done by simple averaging of the elemental nodal linear momentum arising at the node from the connected cells:
\begin{equation}
    \boldsymbol{{p}}_a = \frac{1}{\Lambda_a^e} \sum _{ e \in \Lambda_a^e} \boldsymbol{p}_{ea},
\end{equation}
where $\Lambda_a^e$ is the number of cells connected to a node $a$.

\textit{Step five, interpolate filtering linear momentum:} This is the final step where we obtain the filtering linear momentum 
\begin{equation}
    \boldsymbol{\bar{p}}_f = \frac{1}{\Lambda_f^a} \sum _{ a \in \Lambda_f^a} \boldsymbol{p}_{a},
\end{equation}
where $\Lambda_f^a$ is the number of nodes connected to a face $f$. These steps ensure that the deformation gradient is updated through a continuous field and ensure the compatibility requirements. These steps are schematically represented in \cite{haider2017first}. 

\subsection{Discreet angular momentum projection algorithm}
To satisfy the angular momentum conservation law, a projection-based method was developed by Haider et al. \cite{haider2017first}, following the work done by Aguirre et al. \cite{aguirre2014vertex}.In this approach, the right-hand side of the semi-discrete linear momentum equation, Eq.\ref{eq : semi-discret linear momentum}, represented as $\boldsymbol{\Dot{p}}$, is adjusted to maintain the conservation of angular momentum while simultaneously preserving the conservation of total linear momentum. This modification is expressed as:
\begin{equation}
    \boldsymbol{\Dot{p}}_e = \frac{\partial \boldsymbol{p}}{\partial t} = \frac{1}{\Omega_e }  \sum_{ f \in \Lambda_e^f} (\boldsymbol{{P}} \boldsymbol{N} \, \Delta S)_f .
\end{equation}
Following a Lagrange Multiplier minimization procedure, the corrected right-hand side of the linear momentum equation is modified as,
\begin{equation}
    \boldsymbol{\bar{\Dot{p}}}_e = \boldsymbol{\Dot{p}}_e + \boldsymbol{\lambda_{Ang}} \times \boldsymbol{\mathcal{X}}_e +\boldsymbol{\lambda_{Lin}}.
\end{equation}
here, $\{\boldsymbol{\lambda_{Ang}},\boldsymbol{\lambda_{Lin}}\}$ are the Lagrange Multipliers given by :
\begin{multline}
    \left[\begin{array}{c}
\boldsymbol{\lambda}_{\text {Ang }} \\
\boldsymbol{\lambda}_{\text {Lin }}
\end{array}\right]=\left[\begin{array}{cc}
\sum_e \Omega_e\left[\boldsymbol{\mathcal{X}}_e \otimes \boldsymbol{\mathcal { X }}_e-\left(\boldsymbol{\mathcal { X }}_e \cdot \boldsymbol{\mathcal { X }}_e\right) \boldsymbol{I}\right] & -\sum_e \Omega^e \hat{\boldsymbol{\mathcal {X }}}_e \\
\sum_e \Omega_e \hat{\boldsymbol{\mathcal { X }}}_e & -\sum_e \Omega^e
\end{array}\right]^{-1}\\
\left[\begin{array}{c}
\sum_e \Omega_e\left(\boldsymbol{\mathcal { X }}_e \times \dot{\boldsymbol{p}}_e\right)-\boldsymbol{T}_{\text {tor }}^e \\
\sum_e\left(\Omega_e \dot{\boldsymbol{p}}_e-\boldsymbol{T}_{\text {tor }}^e\right)
\end{array}\right]
\end{multline}
where $\boldsymbol{T}_{{tor }}^e$ is the external torque, $\boldsymbol{\mathcal{X}}_e$ represents the cell center position vector in a consistent manner with each RK stage,
\begin{equation}
    \boldsymbol{\mathcal { X }}_e= \begin{cases}\boldsymbol{x}_e^n, & RK stage=1 \\ \boldsymbol{x}_e^{n+1 / 2}+\frac{\Delta t}{2 \rho_0} \boldsymbol{p}_e^{(1)}, & RK stage=2
    \end{cases}
\end{equation}
and,
\begin{equation}
    \boldsymbol{x}_e^{n+1 / 2}=\boldsymbol{x}_e^n+\frac{\Delta t}{2 \rho_0} \boldsymbol{p}_e^n, \quad \left[\hat{\mathcal{X}}_e\right]_{i k}=\mathcal{E}_{i j k}\left[\mathcal{X}_e\right]_j.
\end{equation}

	\subsection{A complete algorithm}
	A complete solution algorithm is described in Algorithm.\ref{alg: algorithm 1}.
	\begin{algorithm}[h] 
		\caption{Time update of conservation variables}
		\begin{algorithmic}[1]
			\Require $\boldsymbol{\mathcal{U}}_e^n$
			\Ensure $\boldsymbol{\mathcal{U}}_e^{n+1}$
			\State Calculate time step: $\Delta t^n $
			\State Store conservation variables: $\mathbf{U}_e^{\text{old}} = \mathbf{U}_e^n$
			\For{Runge Kutta stage $= 1 \text{ to } 2$}
			\State Evaluate wave speeds: $U_P, \ U_S $
			\State Compute PK1 stresses: $\mathbf{P}_e$
			\State Apply the linear reconstruction procedure:
			\State \quad Obtain reconstructed values at faces: $\mathbf{U}_f^L, \ \mathbf{U}_f^R $
			\State Apply Roe-type Riemann solver:
			\State \quad Rotate the global frame to the local frame of reference for each face.
			\State \quad Find the fluxes at the interior faces.
			\State \quad Rotate the fluxes back to the global frame
			\State \quad Apply boundary conditions.
			\State Apply curl-free algorithm:
			\State \quad Find the linear momentum at each face.
			\State \quad Obtain corrected linear momentum.
			\State Apply the angular momentum projection algorithm.
			\State Solve governing equations: $\mathbf{U}_e = \mathbf{U}_e + \Delta t^n \dot{\mathbf{U}}_e$
			\EndFor
			\State Update conservation variables: $\mathbf{U}_e^{n+1} = \frac{1}{2} (\mathbf{U}_e + \mathbf{U}_e^{\text{old}})$
		\end{algorithmic}\label{alg: algorithm 1}
	\end{algorithm}

%% file: sections/test_cases/numerical_test_cases.tex
\section{Numerical Test Cases}
\label{sec:Numerical Test Cases}
This section presents a set of benchmark problems designed to evaluate the accuracy and robustness of the proposed computational framework. The newly developed Roe-type Riemann solver has been implemented within the OpenFOAM v7 environment by extending the \verb|explicitSolidDynamics| toolkit \cite{haider2019toolkit}.

The first benchmark investigates wave propagation phenomena, including a three-dimensional pile-driving case involving shockwave transmission through a solid medium. The second benchmark addresses stress distribution in a three-dimensional narrow structural component with a T-shaped cross-section. The final test case examines dynamic deformation through the bending response of a two-dimensional beam under both static and dynamic loading conditions. The cases are compared against analytical solutions and the numerical solutions. The reference numerical solutions are obtained using the \verb|solids4foam| toolkit \cite{cardiff2018open} utilizing the a second-order displacement-based formulation model named \verb|linearGeometryTotalDisplacement|.

For steady-state simulations, appropriate damping is introduced as a source term in the governing equations to ensure convergence and physical realism. Such damping may arise from microstructural imperfections, interfacial friction, or intermittent contact at joints. In this study, damping is modeled  by a linear viscous damper \cite{tsui2013finite}, represented by the relation :
\begin{equation}
    \boldsymbol{b} = c \,\boldsymbol{p}.
\end{equation}
where $c$ denotes the damping coefficient and is evaluated based on  Bernoulli-Euler beam theory\cite{rao2019vibration}. The magnitude of damping is chosen based on the underlying physical characteristics of each problem.

\input{sections/test_cases/pile_driving/pile_driving}
\input{sections/test_cases/narrow_T_section/narrow_T_section}

\input{sections/test_cases/beam_bending/beam_bending}

%% file: sections/test_cases/pile_driving/pile_driving.tex
\subsection{Wave Propagation: Pile Driving Benchmark}
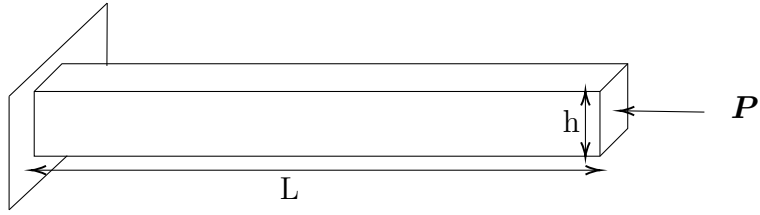
\begin{figure}[t]
    \centering

\begin{tikzpicture}[x=0.5pt,y=0.5pt,yscale=-1,xscale=1]

\draw    (607,142.6) -- (545,141.63) ;
\draw [shift={(543,141.6)}, rotate = 0.9] [color={rgb, 255:red, 0; green, 0; blue, 0 }  ][line width=0.75]    (10.93,-3.29) .. controls (6.95,-1.4) and (3.31,-0.3) .. (0,0) .. controls (3.31,0.3) and (6.95,1.4) .. (10.93,3.29)   ;
\draw   (100,127) -- (121,106) -- (549,106) -- (549,155) -- (528,176) -- (100,176) -- cycle ; \draw   (549,106) -- (528,127) -- (100,127) ; \draw   (528,127) -- (528,176) ;
\draw    (81,130.6) -- (155.24,60.07) ;
\draw    (81,130.6) -- (81,216.6) ;
\draw    (81,216.6) -- (125,175.6) ;
\draw    (100,186.6) -- (526,186.6) ;
\draw [shift={(528,186.6)}, rotate = 180] [color={rgb, 255:red, 0; green, 0; blue, 0 }  ][line width=0.75]    (10.93,-3.29) .. controls (6.95,-1.4) and (3.31,-0.3) .. (0,0) .. controls (3.31,0.3) and (6.95,1.4) .. (10.93,3.29)   ;
\draw [shift={(98,186.6)}, rotate = 0] [color={rgb, 255:red, 0; green, 0; blue, 0 }  ][line width=0.75]    (10.93,-3.29) .. controls (6.95,-1.4) and (3.31,-0.3) .. (0,0) .. controls (3.31,0.3) and (6.95,1.4) .. (10.93,3.29)   ;
\draw    (517,174.4) -- (517,128.6) ;
\draw [shift={(517,126.6)}, rotate = 90] [color={rgb, 255:red, 0; green, 0; blue, 0 }  ][line width=0.75]    (10.93,-3.29) .. controls (6.95,-1.4) and (3.31,-0.3) .. (0,0) .. controls (3.31,0.3) and (6.95,1.4) .. (10.93,3.29)   ;
\draw [shift={(517,176.4)}, rotate = 270] [color={rgb, 255:red, 0; green, 0; blue, 0 }  ][line width=0.75]    (10.93,-3.29) .. controls (6.95,-1.4) and (3.31,-0.3) .. (0,0) .. controls (3.31,0.3) and (6.95,1.4) .. (10.93,3.29)   ;
\draw    (155.24,60.07) -- (155,107.6) ;

\draw (625.08,128.52) node [anchor=north west][inner sep=0.75pt]    {$\boldsymbol{P}$};
\draw (284,190) node [anchor=north west][inner sep=0.75pt]   [align=left] {L};
\draw (498,139) node [anchor=north west][inner sep=0.75pt]   [align=left] {h};

\end{tikzpicture}

    \caption{Pile driving test case - geometry}
    \label{fig:Pile-geometry}
\end{figure}

To evaluate the accuracy of the proposed numerical framework in capturing stress wave propagation in a linear elastic medium, we consider a classical pile driving benchmark problem. The setup involves a prismatic bar (pile) of length \( L = 10 \, m \), rigidly fixed at one end and subjected to a step compressive load at the other, as shown in Figure~\ref{fig:Pile-geometry}. This problem has been widely studied both analytically \cite{clough1979dynamic} and numerically using the finite volume method (FVM) \cite{lee2013development, breil20203d}.
		
The pile has a unit cross-sectional area and is discretized using a three-dimensional grid of \( (100 \times 10 \times 10) \) cells. The material properties,  density $\rho = 8000$ kg/m$^3$, Young’s modulus 
$E=200\times 10^9$ Pa, and Poisson's ratio $\nu = 0$, yield a wave speed of \( U_p = 5000 \, m/s \). A Courant–Friedrichs–Lewy (CFL) number of 0.5 is employed for the first- and second-order finite volume methods (FVMs).

\begin{figure}[H]
    \centering
    \includegraphics[scale = 0.7]{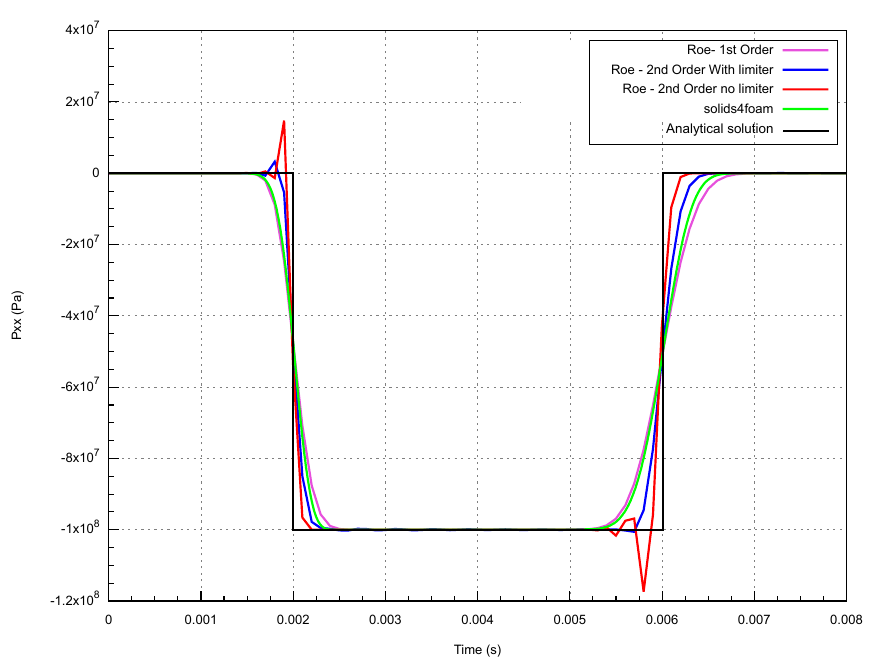}
    \caption{Pile driving test case - Stress history at fixed end, x = 0}
    \label{fig:Pile-shock}
\end{figure}

The applied boundary loading is given by:
\begin{equation}  
	P(L,t)=\left\{\begin{array}{c c}  
		{0}, & {t < 0} \\  
		{-P}, & {t \geq 0}  
	\end{array}\right., \qquad P_0 = 5 \times 10^7 \text{ N}. 
	\label{eq:pile load}
\end{equation}
This initiates a compressive stress wave of magnitude \( -5 \times 10^7 \) Pa that travels toward the fixed boundary at $x=0$. Upon reflection, the wave reverses direction and doubles in amplitude due to the boundary condition, as illustrated in Figure~\ref{fig:Pile-shock}.

Figure~\ref{fig:Pile-shock} compares the numerical results obtained using the first- and second-order FVMs. The first-order scheme exhibits pronounced numerical diffusion, resulting in smeared wave fronts and inaccuracies in long-term wave propagation. In contrast, the second-order method significantly improves resolution and accurately captures the propagation and reflection of stress waves. Moreover, using a limiter in the second-order scheme effectively reduces spurious oscillations near sharp gradients, further enhancing the solution quality. The second-order solution generated by the proposed solver demonstrates enhanced accuracy relative to the results obtained using the solids4foam framework.

%% file: sections/test_cases/narrow_T_section/narrow_T_section.tex
\subsection{Stress Analysis: Narrow T-section component under tension}
\begin{figure}[H]
    \centering
    \includegraphics[scale = 0.35]{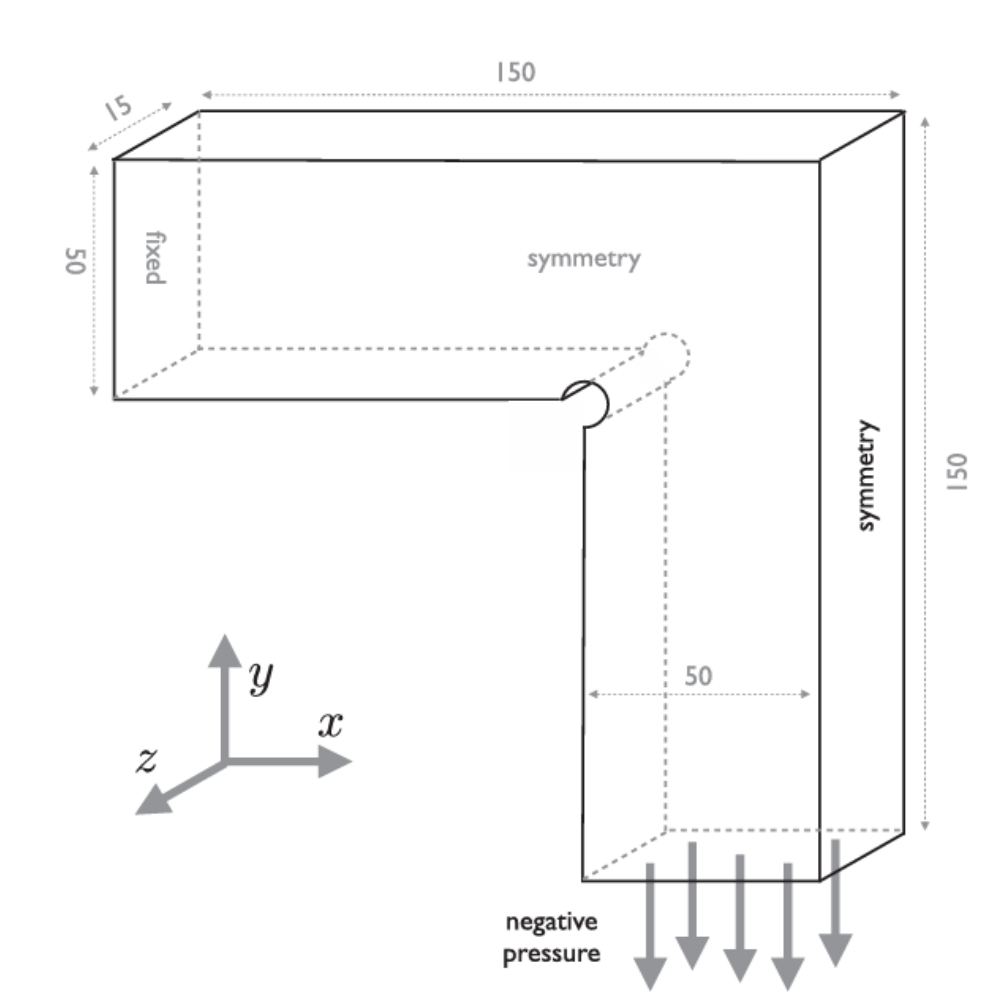}
    \caption{ T member geometry}
    \label{fig: T member geomtry}
\end{figure}

\begin{figure}[H]
    \centering
    \includegraphics[scale = 0.7]{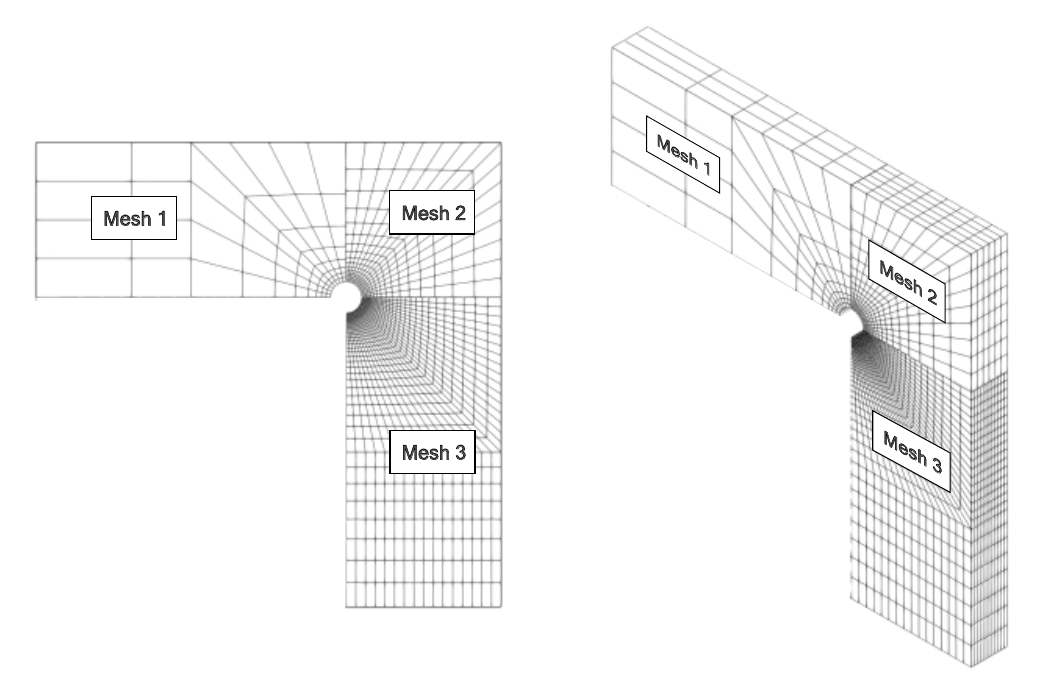}
    \caption{ T member mesh}
    \label{fig: T member mesh}
\end{figure}

The analyzed test case involves a slender engineering component with a T-shaped cross-section Figure~\ref{fig: T member geomtry}, designed based on established benchmarks in the literature \cite{cardiff2016block,demirdvzic1997benchmark}. The geometry of the T-section includes a circular hole with a radius of $5$ mm introduced at a location anticipated to exhibit stress concentration. Only one-quarter of the component is considered in the computational domain due to symmetry. The lower surface of the component is subjected to a constant negative pressure of 1 MPa, while the upper left surface is completely constrained. The material properties,  density $\rho = 7600$ kg/m$^3$, Young’s modulus 
$E=210\times 10^9$ Pa, and Poisson's ratio $\nu = 0.3$.
Three hexahedral meshes are used to evaluate the stress distribution. The coarsest mesh has 624 cells, while the finest mesh has 319488 cells. Figure~\ref{fig: T member mesh}.

\begin{figure}[H]
    \centering
    \includegraphics[width=1\linewidth]{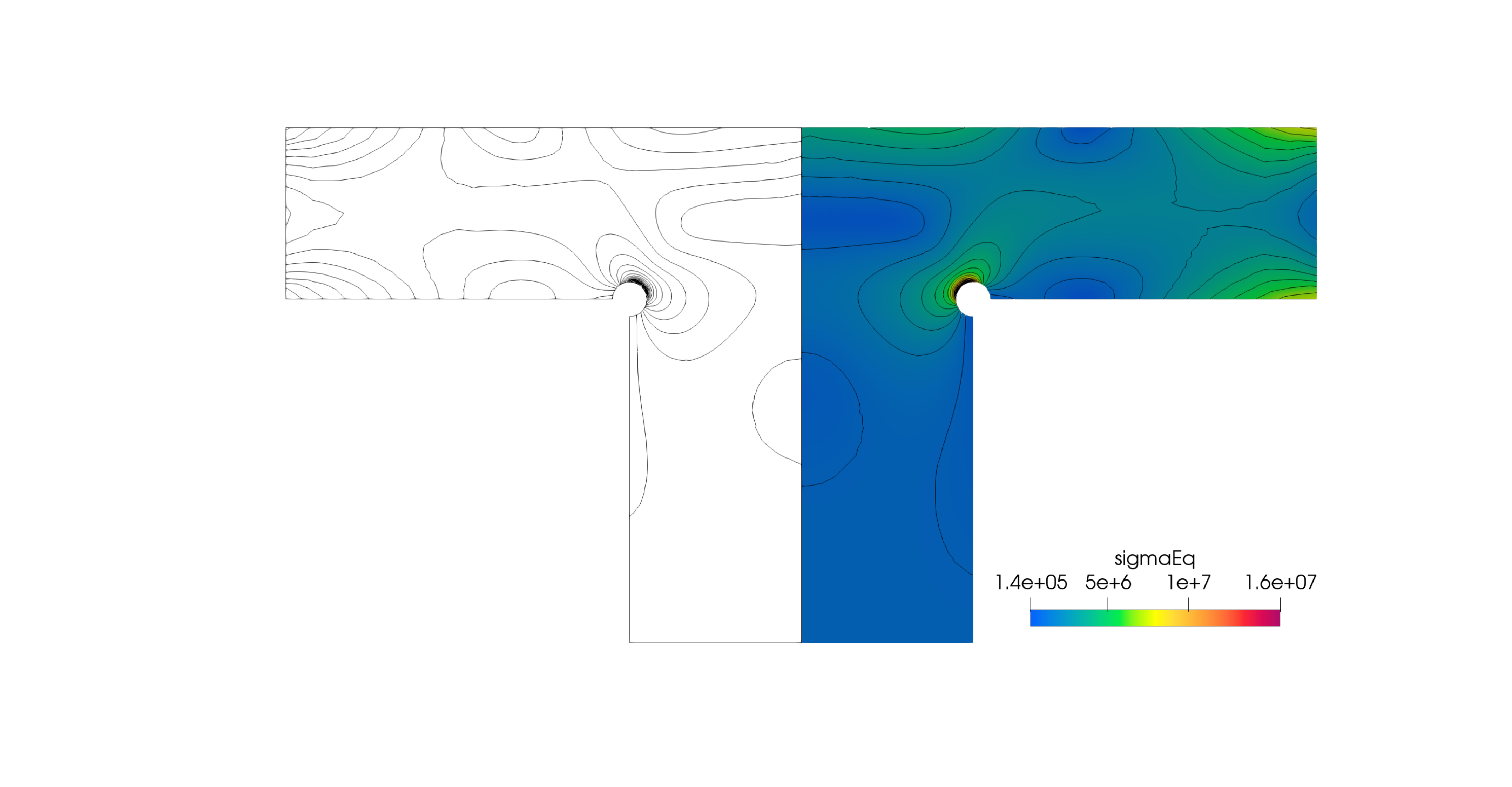}
    \caption{Contour plot for the stress component in XX}
    \label{fig: T member contour}
\end{figure}

\begin{figure}[H]
    \centering
    \includegraphics[width=0.8\linewidth]{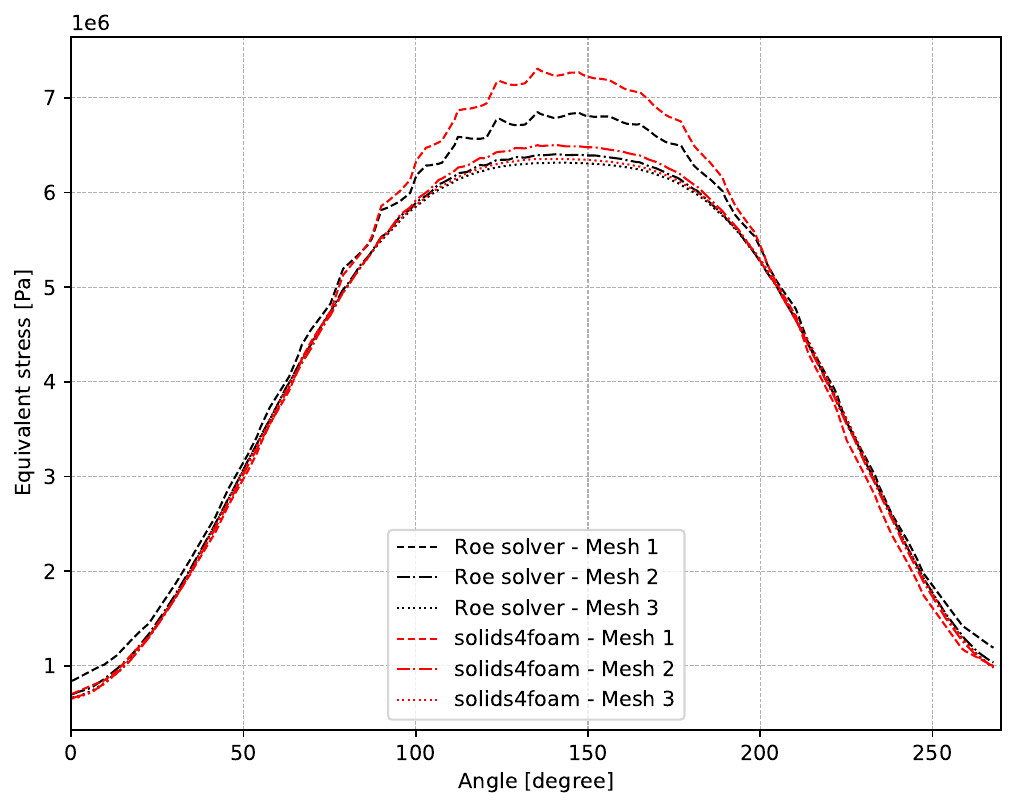}
    \caption{ Narrow T-section component: equivalent stress plotted along z=0 plane, r = 1.5R.}
    \label{fig: T member equivalent stress plotted along z=0 plane, r = 1.5R}
\end{figure}

Figure~\ref{fig: T member contour} compares the contour lines of the equivalent stress resulting from the solution of the solids4foam toolkit using the linearGeometryTotalDisplacement solid model on the left and the current approach on the right. On the other hand, Figure~\ref{fig: T member equivalent stress plotted along z=0 plane, r = 1.5R} presents the same comparison but along a profile position at \(r = 1.5R\) in the plane where \(z = 0\). The results show rapid convergence towards a grid-independent solution across all cases, with agreement between the algorithms. It can also be seen that equivalent stress profiles in Figure~\ref{fig: T member equivalent stress plotted along z=0 plane, r = 1.5R} obtained by the proposed solver are significantly more accurate on coarser meshes.

%% file: sections/test_cases/beam_bending/beam_bending.tex
\subsection{Dynamic Deformation:  Deformation of a Cantilever Beam}
\begin{figure}[H]
    \centering
    \includegraphics[scale = 1]{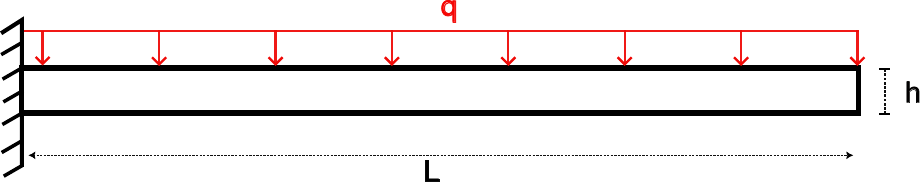}
    \caption{Cantilever beam under distributed load}
    \label{fig:cantilever}
\end{figure}

This computational analysis centers around the deformation behavior of a fixed-free cantilever structure that supports external loads. We chose this benchmark specifically for future research work involving fluid-solid interaction problems. Specifically, the benchmark represents the validation and verification of the solid component deformations under blast loading \cite{bailoor2017fluid,pasquariello2016cut,ning2021novel} or resembling hydrodynamic instabilities observed in segmented rocket motors \cite{giordano2005shock}.
The structural configuration is illustrated in Figure~\ref{fig:cantilever}, where the beam is fixed at a single end and subjected to distributed loads on its top surface. it has a length of $(50)$ mm and hight of $(1)$ mm The load resembles a pressure difference equal to $(1$ atm $\approx 1 e+5$ Pa).

Following the classical beam theory relations of a cantilever plate \cite{fenner1991mechanics}, the displacement at the free end is given by:
\begin{equation}
    dy =\frac{q L^3}{8EI}
\end{equation}
where, $dy$ is the displacement at the free end, $q$ is the uniformly distributed load, $L$ is the length of the cantilever, $h$ is its hight, $E$ is Young's modulus, and $I$ represents the moment of inertia of the cross-section. The material properties,  density $\rho = 7600$ kg/m$^3$, Young’s modulus $E=220\times 10^9$ Pa, and Poisson's ratio $\nu = 0$. The resulting analytical displacement is $dy=4.26$.

\begin{figure}[H]
    \centering
    \includegraphics[width=0.9\linewidth]{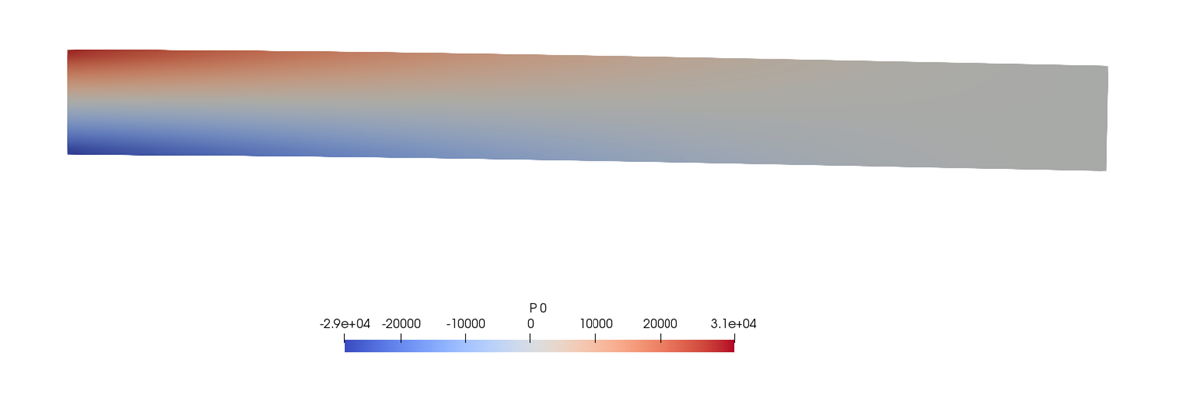}
    \caption{Beam bending: $P_{11}$ distribution }
    \label{fig:beamBending_staticStress}
\end{figure}

Three levels of grid spacing are employed for calculations. The resulting displacements at the middle point of the free end face are recorded, and the error compared with the theoretical value is presented in Table~\ref{table:beamBending_error}. The stress distribution is plotted in Fig.\ref{fig:beamBending_staticStress}.   Displacement error comparisons reveal that the proposed Roe-type Riemann solver consistently produces significantly more accurate results across all test cases.

\begin{table}[H]
\centering
\begin{tabularx}{\textwidth}{llXXX}
\hline
                        & Mesh Size  & 50x3  & 100x4 & 200x8 \\ \hline
\multirow{2}{*}{Roe solver}    & displacement & 4.10  & 4.19  & 4.27  \\
                        & Error      & 3.722 & 1.67  & 0.2 \\ \hline
\multirow{2}{*}{solids4foam} & displacement & 3.02  & 3.64  & 3.99  \\
                        & Error      & 29.1  & 14.4  & 6.24  \\ \hline
\end{tabularx}
\caption{Beam Bending: Maximum displacement (mm) and relative error(\%) compared between the developed Roe-type Riemann solver and solids4foam solver using linearGeometryTotalDisplacement solid model}
\label{table:beamBending_error}
\end{table}

%% file: sections/conclusion.tex
\section{Conclusion}
\label{sec:Conclusion}
This paper introduces a novel finite volume framework for solid dynamics based on a first-order momentum–deformation formulation. Central to this approach is the development of a new Roe-type Riemann solver specifically tailored for solid mechanics, leveraging the system's hyperbolicity and rotational invariance to accurately capture wave propagation and dynamic deformation phenomena.

The method has been successfully implemented within the OpenFOAM platform, enabling scalable and practical applications. Its performance has been validated through a series of two- and three-dimensional benchmark problems in linear elasticity, including shock wave propagation and transient stress analysis. The results of the proposed Roe-type Riemann solver show excellent agreement with analytical solutions and exhibit significantly improved accuracy compared to reference results obtained using a displacement-based formulation.

By formulating the problem in terms of first-order hyperbolic conservation laws, the framework benefits from improved numerical stability and consistency, particularly suited for wave-dominated and shock-sensitive regimes. Moreover, its modular structure and compatibility with upwind discretization schemes make it a promising candidate for extension to nonlinear solid mechanics and coupled fluid–structure interaction (FSI) problems.

Overall, this work establishes a solid foundation for advancing finite volume methods in solid dynamics. Future research will focus on extending the proposed solver to handle material and geometric nonlinearities, large deformations, and multi-physics interactions, further broadening its applicability in industrial, aerospace, and biomechanical engineering domains.